\documentclass[aip,pra,showpacs,reprint,amsmath,amssymb,superscriptaddress,floatfix]{revtex4-1}

\usepackage{natbib}

\usepackage[utf8]{inputenc}
\usepackage{graphicx}
\usepackage{dcolumn}
\usepackage{bm}
\usepackage{array}

\usepackage{epstopdf}
\usepackage{multirow}
\usepackage{color}
\usepackage{enumerate}
\usepackage{makecell}

\begin{document}

\title{Dynamics of superconducting qubit relaxation times}

\author{M. Carroll \textsuperscript{*}\thanks{\noindent\textsuperscript{*} These authors contributed equally}}
\email{malcolm.carroll@ibm.com}
\affiliation{IBM Quantum, IBM T.J. Watson Research Center,Yorktown Heights, NY, 10598, USA}
\author{S. Rosenblatt\textsuperscript{*}\thanks{\noindent\textsuperscript{*} These authors contributed equally}}
\email{srosenb@us.ibm.com}
\affiliation{IBM Quantum, IBM T.J. Watson Research Center,Yorktown Heights, NY, 10598, USA}
\author{P. Jurcevic} 
\affiliation{IBM Quantum, IBM T.J. Watson Research Center,Yorktown Heights, NY, 10598, USA}
\author{I. Lauer} 
\affiliation{IBM Quantum, IBM T.J. Watson Research Center,Yorktown Heights, NY, 10598, USA}
\author{A. Kandala} 
\email{akandala@us.ibm.com}
\affiliation{IBM Quantum, IBM T.J. Watson Research Center,Yorktown Heights, NY, 10598, USA}

\begin{abstract}
Superconducting qubits are a leading candidate for quantum computing but display temporal fluctuations in their energy relaxation times $T_1$. This introduces instabilities in multi-qubit device performance. Furthermore, autocorrelation in these time fluctuations introduces challenges for obtaining representative measures of $T_1$ for process optimization and device screening. These $T_1$ fluctuations are often attributed to time varying coupling of the qubit to defects, putative two level systems (TLSs). In this work, we develop a technique to probe the spectral and temporal dynamics of $T_1$ in single junction transmons by repeated $T_1$ measurements in the frequency vicinity of the bare qubit transition, via the AC-Stark effect. Across 10 qubits, we observe strong correlations between the mean $T_1$ averaged over approximately nine months and a snapshot of an equally weighted $T_1$ average over the Stark shifted frequency range. These observations are suggestive of an ergodic-like spectral diffusion of TLSs dominating $T_1$, and offer a promising path to more rapid $T_1$ characterization for device screening and process optimization.

\end{abstract}

\date{\today}
\maketitle
\section{Introduction}
\label{Introduction}
Superconducting qubits are a leading platform for quantum computing~\cite{zhang2020,arute2019}. This has been driven, in part, by improvements in coherence times over five orders of magnitude since the realization of coherent dynamics in a cooper pair box~\cite{nakamura1999}. However, further improving coherence times remains crucial for enhancing the scope of noisy superconducting quantum processors as well as the long term challenge of building a fault tolerant quantum computer. Recent advances~\cite{hong2020,kandala2020,hashim2020,Foxen2020} in two-qubit gate control have placed their fidelities at the cusp of their coherence limit, implying that improvements in coherence could directly drive gate fidelities past the fault tolerant threshold. In this context, coherence stability and its impact on multi-qubit device performance is also an important theme, since superconducting qubits have been shown to display large {\color{black}and correlated temporal fluctuations (i.e., $1/f^\alpha $)} in their energy relaxation times $T_1$\cite{muller2015,Paladino_RevModPhys.86.361,Weissman_RevModPhys.60.537,klimov_fluctuations_2018,kandala2019,burnett2019,schlor2019}. This places additional challenges for benchmarking the coherence properties of these devices~\cite{burnett2019}, 
and also for error mitigation strategies such as zero noise extrapolation~\cite{kandala2019}.  

The fluctuations of qubit $T_1$ are often attributed to resonant couplings with two level systems (TLSs) that have been historically studied in the context of amorphous solids \cite{phillips1972,muller_towards_2019} and their low temperature properties. More recently, TLSs have attracted renewed interest due to their effect on the coherence properties of superconducting quantum circuits\cite{martinis2005,grabovskij2012strain,Barends2013,klimov_fluctuations_2018,lisenfeld2019electric,burnett2019,schlor2019}, and are attributed to defects in amorphous materials at surfaces, interfaces, and the Josephson junction tunnel barrier. Frequency resolved measurements of $T_1$ in flux and stress tunable devices\cite{Barends2013,klimov_fluctuations_2018,lisenfeld2019electric} have also displayed fluctuations, suggesting an environment of TLSs with varying coupling strengths around the qubit frequency. The variability of $T_1$ over time is explained\cite{muller_towards_2019,klimov_fluctuations_2018}, at least in part, by temporal fluctuations in this frequency environment, associated with the spectral diffusion of the TLSs \cite{black_spectral_1977, phillips1972}. 

Furthermore, two-qubit gates that involve frequency excursions~\cite{kandala2020,stehlik2021,klimov_fluctuations_2018} can also interact with TLS near the qubit frequency leading to additional incoherent error. The fluctuations in TLS peak positions, therefore, can also introduce fluctuations in two qubit fidelity. Spectroscopy of defect TLS is, therefore, central to understanding the short and long time $T_1$ and gate fidelity of qubits.

\begin{figure}
    \centering
    \includegraphics[width=85mm, clip,trim = 10.0mm 25.0mm 0.0mm 7.0mm]{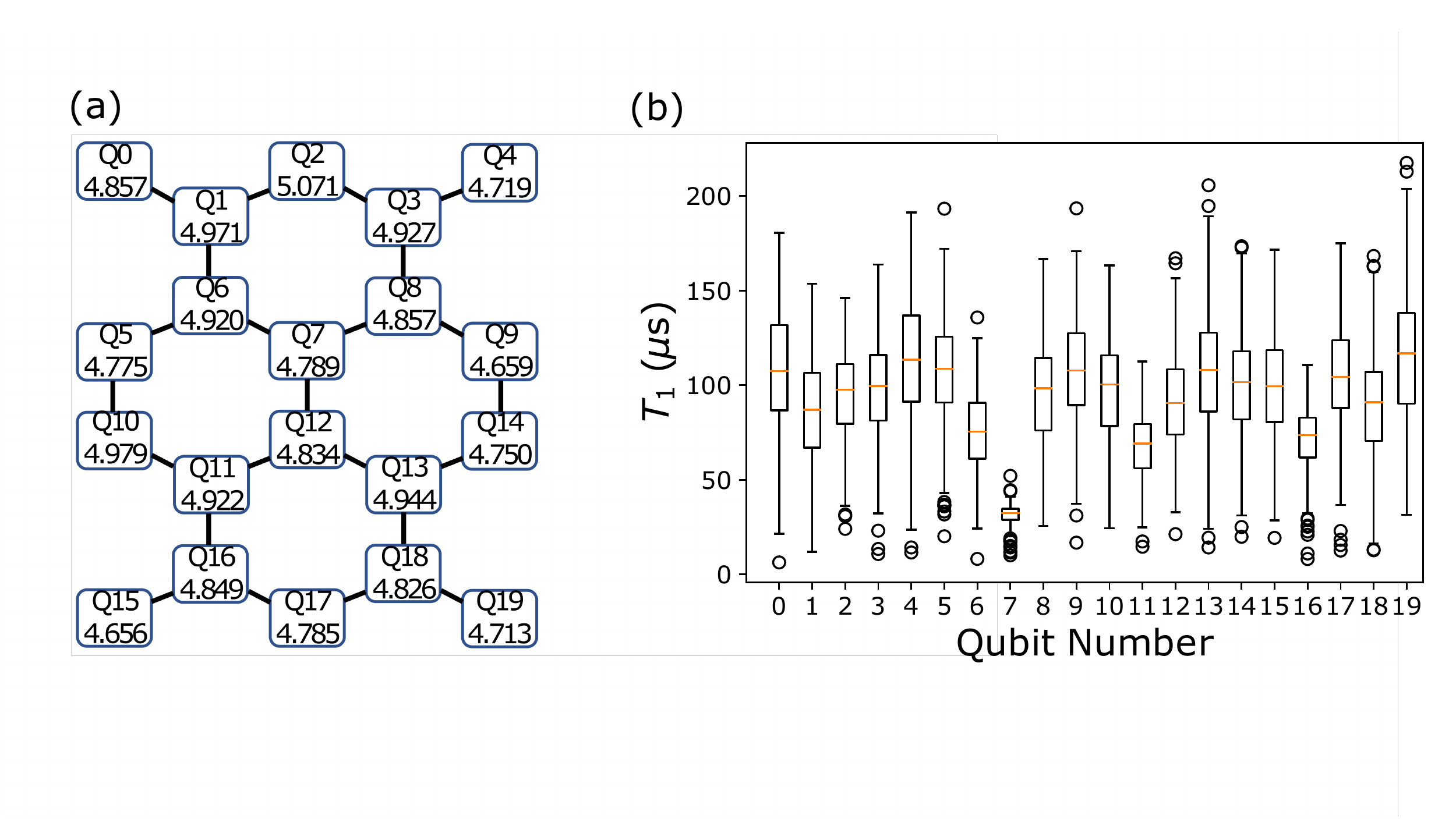}
    \caption{(a) Map of a 20 qubit device including the qubit number, two-qubit connectivity and the qubit frequency in (GHz). (b) Box and whisker plot of $T_1$'s measured daily 
    over approximately nine months. 
    The box covers the first to third quartile, the vertical line indicates the median and the whiskers are drawn to the maximum or minimum values that fall within 1.5 times the interquartile range. All other points are outliers.}
    \label{fig:Figure1}
\end{figure}
Single Josephson junction transmons with fixed frequency couplings represent a successful device architecture achieving networks of over 60 qubits~\cite{zhang2020} with all microwave control and state of the art device coherence. The single junction configuration offers advantages such as reduced sensitivity to flux noise, while preserving the transmon charge insensitivity and also reducing system complexity with fewer control inputs. However, there is little TLS spectroscopy of single junction transmons because of the limited tunability, despite the central importance of understanding the TLS environment both for device and process characterization.

\begin{figure}
    \centering
    \includegraphics[width=8.5 cm, clip,trim = 0.0mm 0.0mm 0.0mm 0.0mm]{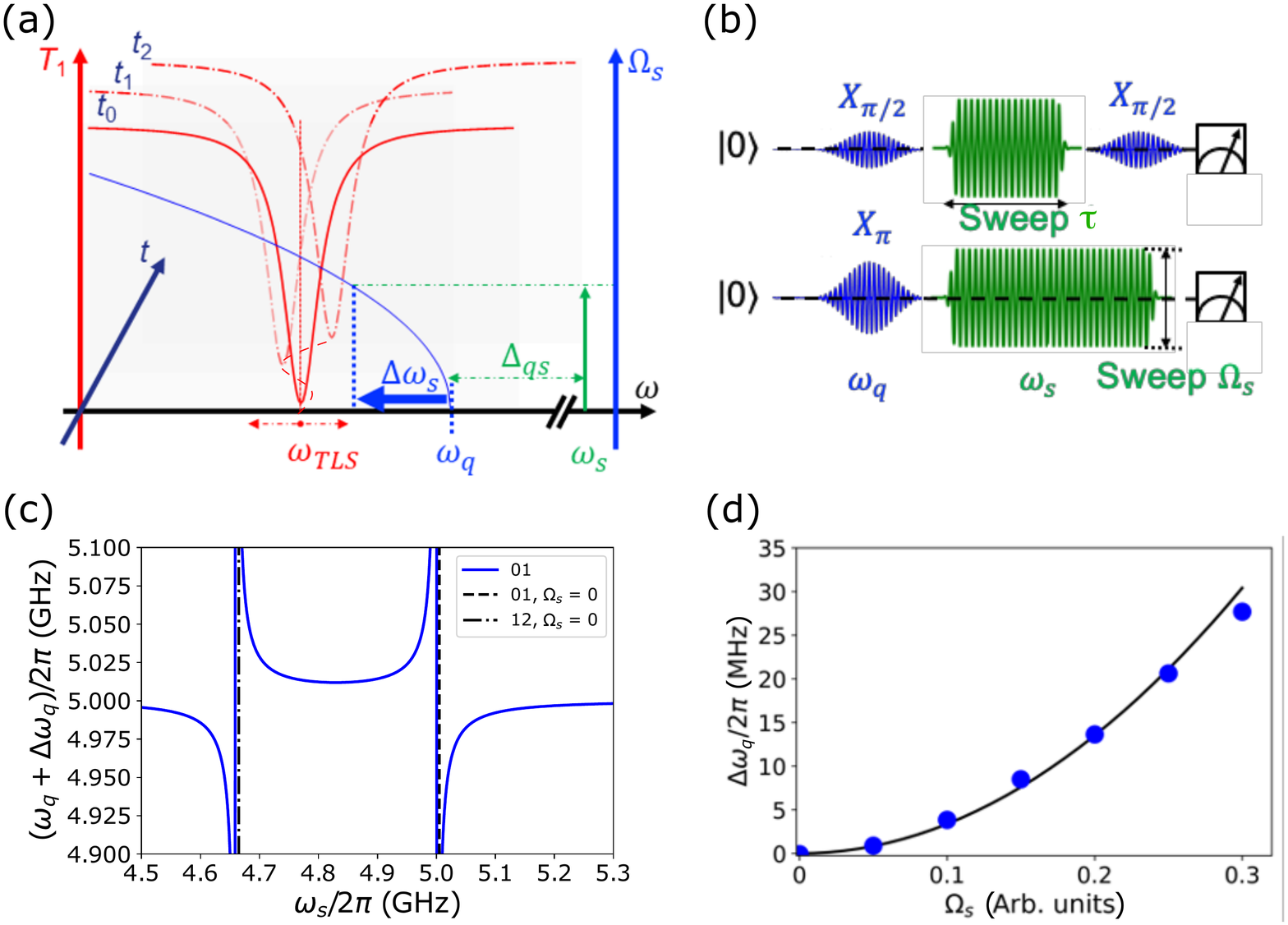}
    \caption{(a) Schematic of qubit $T_1$ response to shift in qubit frequency (red traces), at different times, $t$. The qubit frequency is tuned by an amount $\Delta\omega_q$ by an off-resonant tone placed $|\Delta_{qs}|$ above the qubit frequency $\omega_q$. The dependence of the frequency shift on the off-resonant tone amplitude $\Omega_s$ is depicted by the blue trace. The $T_1$ dips are indicative of the qubit coming into resonance with a strongly coupled TLS, at the frequency $\omega_{TLS}$. The TLS frequency shifts in time due to spectral diffusion, schematically indicated by changes in the $T_1$ dips at different time snapshots. (b) Schematic of a Ramsey pulse sequence used to calibrate $\Delta \omega_q$ as a function of $\Omega_s$ (top); and (bottom) schematic pulse sequence for the relaxation time spectroscopy. For each $\Omega_s$ (i.e., $\Delta \omega_q$), the $|1\rangle$ occupation is measured at a fixed time (i.e., $\tau$ = 50 $\mu$s in this work).  (c) {\color{black}An illustrative case of the 01 transition dependence on $\omega_{s}$ for constant $\Omega_s$. The 01 qubit frequency, $\omega_q$ + $\Delta \omega_q$, uses an unperturbed frequency of $\omega_q$ = 5.0 GHz and an anharmonicity of $\delta_{q}$ = -340 MHz. The locations of the unperturbed 01 and 12 transitions are shown as vertical lines overlaid with 5 MHz offsets to make their locations more visible on the figure. Negative and positive qubit shifts can be produced and large shifts can be induced depending on $\Delta_{qs}$.} (d) Measured $\Delta\omega_q$ as a function of normalized DAC amplitude, $\Omega_s$ using the AC Stark shifted Ramsey technique. Solid line is a quadratic fit functionally consistent with a perturbative model.}

    \label{fig:Figure2}
\end{figure}

In this work, we introduce an all-microwave technique for the fast spectroscopy of TLSs in single junction transmon qubits that requires no additional hardware resources.  In contrast to flux based approaches to TLS spectroscopy, we employ off-resonant microwave tones to drive AC-Stark shifts of the fundamental qubit transition and spectrally resolve qubit relaxation times. Dips in relaxation times serve as a probe of the frequency location of a strongly coupled TLS. We use repeated frequency sweeps to probe the time dynamics of the relaxation probabilities including tracking the spectral diffusion of strongly coupled TLS. 
Across 10 qubits, we observe strong correlations between the long time mean, averaged over several months $\langle{T_1}\rangle_{T}$, and the short time mean, averaged around the local qubit frequency $\langle{T_1}\rangle_{\omega,t}$.

This strong correlation suggests a quasi-ergodic behavior of the TLS spectral diffusion in the nearby frequency neighborhood of the qubit. In contrast, there is lower correlation between $\langle{T_1}\rangle_{T}$ and $T_1$ measured over a single day. The $\langle{T_1}\rangle_{\omega,t}$ can provide, therefore, a more rapid estimate of long time behavior. 

\section{Device and spectroscopy technique}
\label{section:Technique}
The experiments reported in this letter were performed on ibmq\textunderscore almaden, a 20 qubit processor based off single junction transmons and fixed couplings. The device topology is shown in Fig. \ref{fig:Figure1} (a), and qubit frequencies are around $\sim$5 GHz. Fig. \ref{fig:Figure1} (b) depicts the characteristic spread of the qubit $T_1$s and their mean, from $\sim$250 measurements over 9 months. The 
base plate (to which the device was mounted) temperature of the dilution refrigerator was typically $\sim$13 mK 
excepting several temperature excursions to $\sim$1 K, which were not observed to have any significant effects on the {\color{black} long time $T_1$ time series or distributions of $T_1$ values discussed in this work}, discussed later. 
Several qubits on the device display mean $T_1$s exceeding 100 $\mu$s. However, the large spread in individual qubit $T_1$s highlights the challenge for rapid benchmarking of device coherence, since any single $T_1$ measurement can disagree substantially from its longtime mean. 

We study the spectral dynamics of these $T_1$ times by employing off-resonant microwave tones~\cite{Gambetta2006} to induce an effective frequency shift $\Delta\omega_{q}$ in single junction transmons by the AC Stark effect. 
This has been employed previously for coherent state transfer between coupled qubits that are Stark shifted into resonance~\cite{majer2007}. In this work, shifting the qubit frequency into resonance with a defect TLS induces a faster relaxation time, which in turn is used to detect the frequency location of the TLS \cite{simmonds_decoherence_2004}, as depicted in Fig. \ref{fig:Figure2} (a).
The Stark shift can be described analytically by a Duffing oscillator model \cite{magesan2020, Schneider_PhysRevA.97.062334}
\begin{equation}
    \Delta\omega_{q} = \frac{\delta_{q}\Omega_{s}^{2}}{2\Delta_{qs}(\delta_{q}+\Delta_{qs})}
\label{StarkShift}
\end{equation}
where $\delta_{q}$ is the qubit anharmonicity, $\Omega_{s}$ is the drive amplitude and $\Delta_{qs} = \omega_{q} - \omega_{s}$ is the detuning between the qubit frequency and the Stark tone. 

As seen from the expression above, the magnitude and sign of the Stark shift can be manipulated by the detuning and the drive amplitude of the Stark tone, Fig. \ref{fig:Figure2} (c). Very large frequency shifts can be obtained by driving close to the transmon transitions, but this typically leads to undesired excitations/leakage out the two-state manifold. In this work, we obtain Stark shifts of 10's of MHz, with modest drive amplitudes and a fixed detuning $\Delta_{qs}$ of $\pm$50 MHz. The frequency shifts are experimentally measured using a modified Ramsey sequence \cite{ramsey_molecular_1950}, schematically shown in Fig. \ref{fig:Figure2} (b), and display good agreement with the {\color{black}quadratic dependence of the} perturbative model in the low-drive limit. {\color{black}A representative case is shown in} Fig. \ref{fig:Figure2} (d). 
\begin{figure}[t]
    \centering
    \includegraphics[width=85mm, clip,trim = 0.0mm 0.0mm 0.0mm 0.0mm]{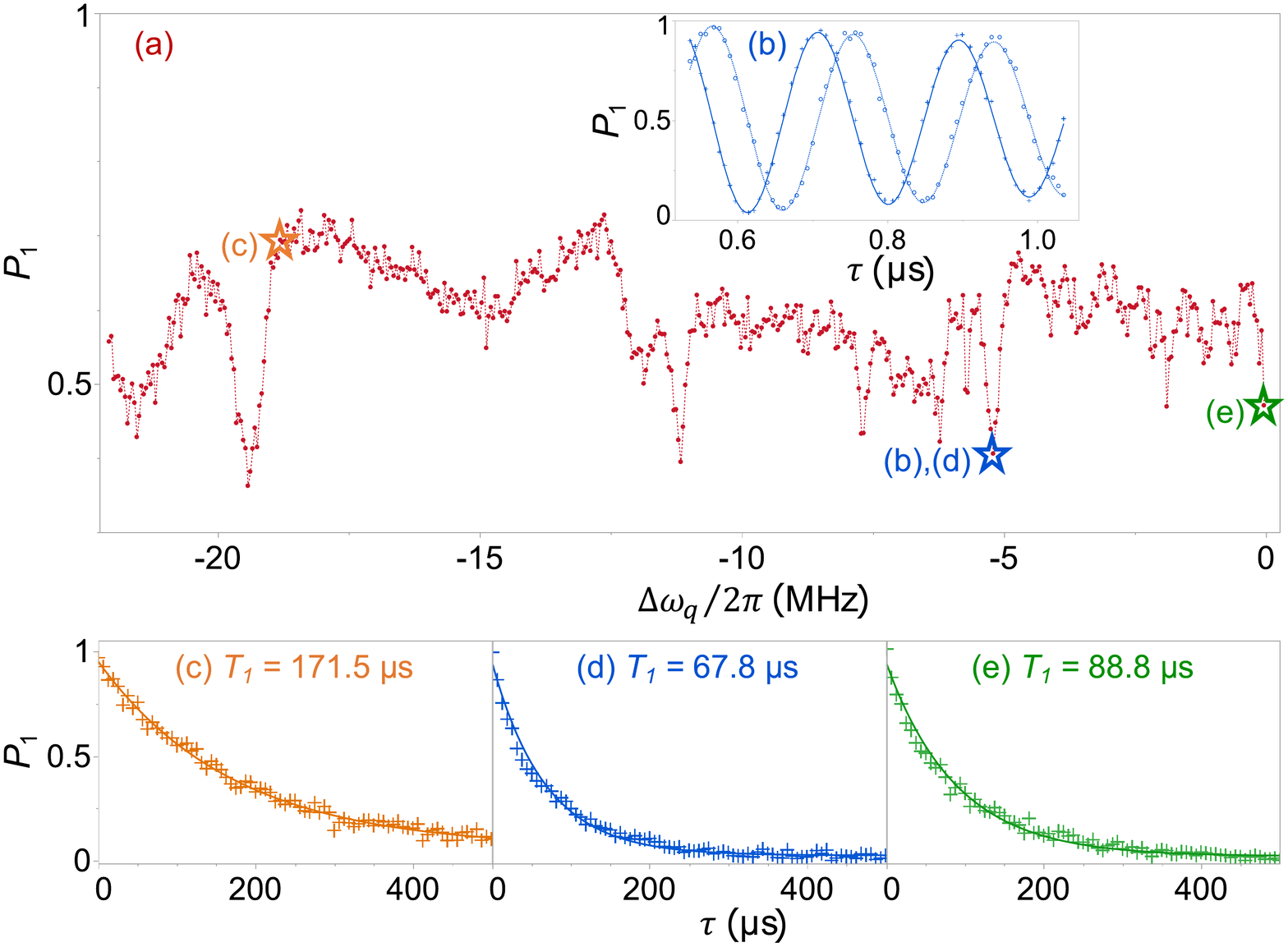}
    \caption{ (a) Measured probability of being in the $|1\rangle$ state, $P_1$, at 50 $\mu$s wait time with varying $\Delta \omega_q$ and tone detuned 50 MHz above ${\omega_{q}/2\pi}$ for $Q_{19}$. (b) Example of Ramsey measurements used to extract frequency shifts ${\Delta\omega_{q}/2\pi}$ from pulse amplitudes. The two curves result from starting the Ramsey oscillations with a $X_{\pi/2}$ or $Y_{\pi/2}$. (c) and (d) are $T_1$ measurement with Stark shifts $\Delta{\omega_{q}/2\pi}$ = -18.9 and -5.3 MHz, respectively. (e) $T_1$ measurement with no Stark shift (i.e., no Stark tone).} 
    \label{fig:Figure3}
\end{figure}

We focus on the spectrally resolved $T_1$ measurements in Fig. \ref{fig:Figure3} that we use as a probe of defect TLS transition frequencies. However, instead of measuring the entire $T_1$ decay, we use the excited state probability, $P_1$, after a fixed delay time as a measure of $T_1$. This speeds up the spectral scans significantly. Our experiments are performed at a repetition rate of 1 kHz, but our scheme can be further accelerated with reset techniques~\cite{egger2018}, which can be crucial for probing faster TLS dynamics. For an effective frequency sweep, we run an amplitude sweep with off-resonant pulses at fixed detuning ($\pm$50 MHz) and duration (delay time of 50 $\mu$s), after exciting the qubit with an initial $\pi$ pulse. The pulsed Stark sequence enables faster spectroscopy by circumventing the need to re-calibrate the $\pi$, $\pi/2$ pulses at every frequency. The off-resonant pulses have Gaussian-square envelopes with a 2$\sigma$ rise-fall profile, where $\sigma$ = 10 ns. This pulse sequence is shown in Fig. \ref{fig:Figure2} (b). The amplitude points in the sweep are then related to Stark shifts by Ramsey sequences. Fig. \ref{fig:Figure3} shows representative data of such a sweep on qubit 19 ($Q_{19}$) with distinctive dips in $P_1$ that we attribute to strongly coupled TLS at their transition frequencies. $T_1$ measurements at Stark tone amplitudes corresponding to high/low $P_1$ points, as seen in the bottom panel of Fig. \ref{fig:Figure3}, explicitly show the substantial variation in $T_1$ as a function of frequency and the consistent tracking of {\color{black}$T_1$ with $P_1$.} 

Variations in $P_1$ can potentially be caused by sources other than TLS. In our experiments, $P_1$ is spectrally resolved to $\sim\pm$25 MHz around the individual qubit frequencies. The narrow frequency range combined with measuring non-neighbor sets of qubits simultaneously avoids strong $P_1$ suppression from resonances with neighboring qubits, the coupling bus or common low-Q parasitic microwave modes. Control experiments show that time insensitive features in the $P_1$ fingerprint are robust to choice of the Stark tone detuning, ruling out a power dependence for the power range used in this work. 
Finally, while a recent report \cite{yan2016} modeled their broadband $T_1$ scatter as arising from quasi-particle fluctuations, this is not sufficient to explain the sharp frequency dependent $P_1$ features depicted, for instance in Fig. \ref{fig:Figure3}. Furthermore, recent experiments on our qubits suggest a quasi-particle limit to $T_1$ that exceeds several milliseconds.\cite{kurter2020}
\section{TLS dynamics and correlations between $P_1(\omega,t)$ and $\langle T_1\rangle_T$}
\label{section:TLS dynamics}

We repeat the line traces of Fig. \ref{fig:Figure3} for both positive and negative 50 MHz detuning, approximately once every 3-4 hours, extended over hundreds of hours for all the qubits. A representative example of the cumulative scans is shown in Fig. \ref{fig:Figure4} for $Q_{15}$. Spectroscopy of the other qubits is shown in the supplemental information, appendix \ref{other spectroscopy}. The TLS dynamics around the qubit frequency are qualitatively similar to previous TLS spectroscopy using flux or stress tunable devices \cite{muller_towards_2019}.

In the case of $Q_{15}$, Fig. \ref{fig:Figure4}, there are prominent dips in relaxation probability around positive 1 MHz, negative 5-10 MHz, and negative 15-20 MHz. The spectral diffusion of the positions of the $T_1$ dips can vary between order of 1 MHz to 10 MHz over the 272 hours of measurement providing a qualitative measure of linewidths. A more quantitative discussion of linewidths can be found in appendix \ref{line widths}. The background is covered by an ensemble of smaller dips of relaxation, Fig. \ref{fig:Figure3}, that also dynamically evolve, with features that are larger than the sampling noise in the measurement.
\begin{figure}[b]
    \centering
    \includegraphics[width=85mm, clip,trim = 0.0mm 2.0mm 0.0mm 0.0mm]{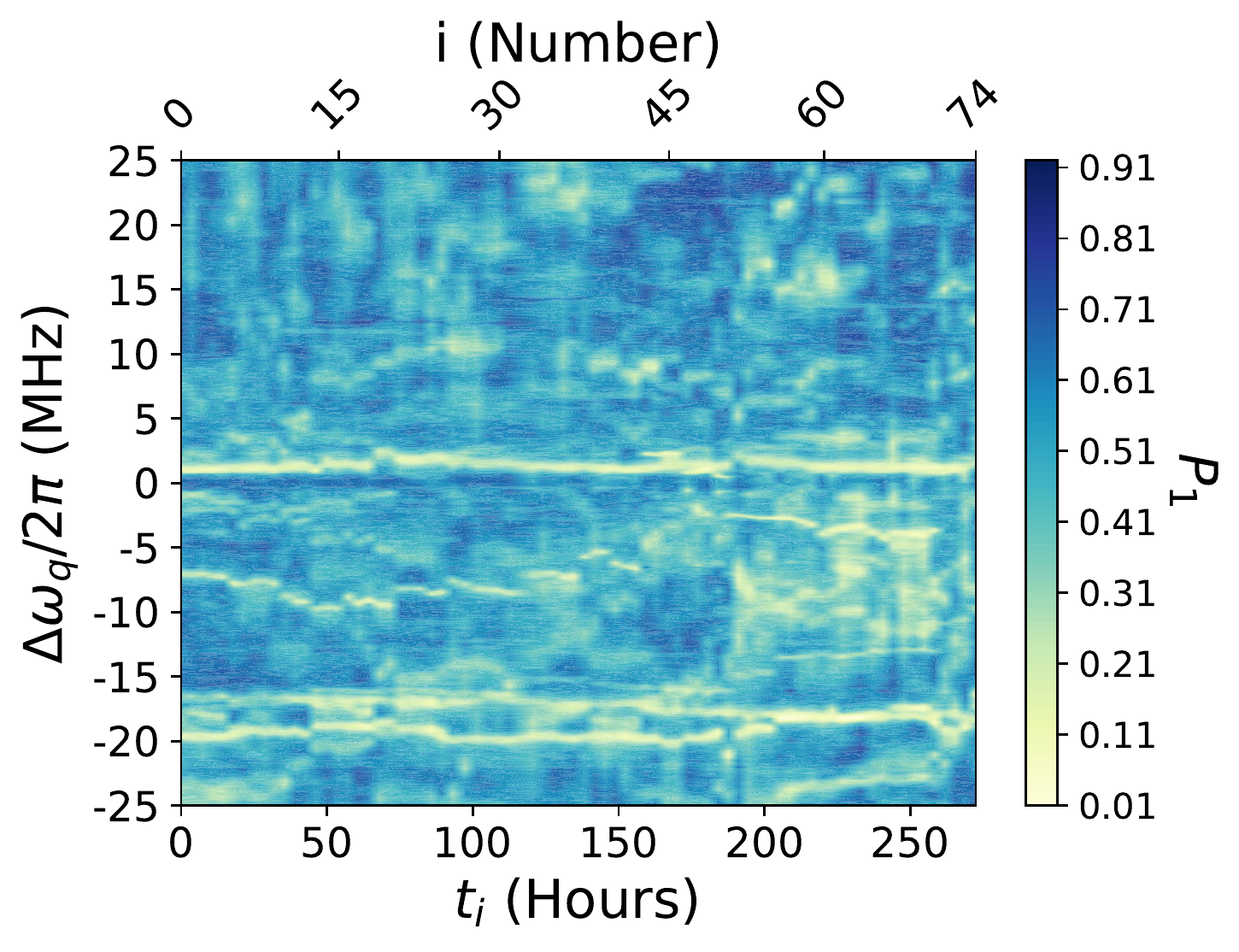}
    \caption{Time dependence of the energy relaxation spectroscopy for $Q_{15}$ using {\color{black}$\Delta\omega_s = \pm$50 MHz and varying $\Omega_{s}$ over 501 different j amplitude points, $\Omega_{s,j}$, to sweep $\Delta\omega_{q}(\Omega_{s,j})$ in a positive or negative direction (see eqn. \ref{StarkShift}).} The $P_1$ is measured at 50 $\mu$s.}
    \label{fig:Figure4}
\end{figure}

As discussed previously, $T_1$ fluctuations introduce uncertainty in the coherence benchmarking, stability of multi-qubit circuit performance and process optimization of superconducting qubit devices. 
In this context of better {\color{black}estimator}, we examine if the long time averages ($T \sim$ 9 months) $\langle T_1 \rangle_{T} $ and $\langle P_1 \rangle_{T} $  are correlated with the frequency neighborhood of the qubit $\langle T_1 \rangle_{\omega,t} $ and $\langle P_1 \rangle_{\omega,t} $, respectively. The averaged relaxation probabilities and $T_1$s are defined as
\begin{equation}
    \langle P_1 \rangle_{T} = \frac{1}{N}\sum_{i=1}^{N}P_1(\omega_q,\tau, T_i)
\label{def4}
\end{equation}

\begin{equation}
    \langle T_1 \rangle_{T} = \frac{1}{N}\sum_{i=1}^{N}T_1(\omega_q, T_i)
\label{def4}
\end{equation}

\begin{equation}
    \langle P_1 \rangle_{\omega,t} = \frac{1}{n}\sum_{i=1}^n \frac{1}{2\Delta\omega}\sum_{\omega_j=-\Delta \omega}^{\Delta \omega}P_1(\omega_q+ \omega_j,\tau,t_i)d\omega_j,
\label{def3}
\end{equation}

{\color{black}
\begin{equation}
    \langle T_1 \rangle_{\omega,t}=\frac{1}{n}\sum_{i=1}^n \frac{1}{2\Delta\omega}\sum_{-\Delta \omega}^{\Delta \omega}\frac{-\tau}{\ln(P_1(\omega_q+ \omega_j,\tau,t_i)d\omega_j)},
\label{def3}
\end{equation}
}
where definitions of variables can be found in table \ref{table:paperdefs}.

\begin{figure*}[t]
    \centering
    \includegraphics[width=\textwidth,height=5cm, clip,trim = 0.0mm 80.0mm 0.0mm 50.0mm]{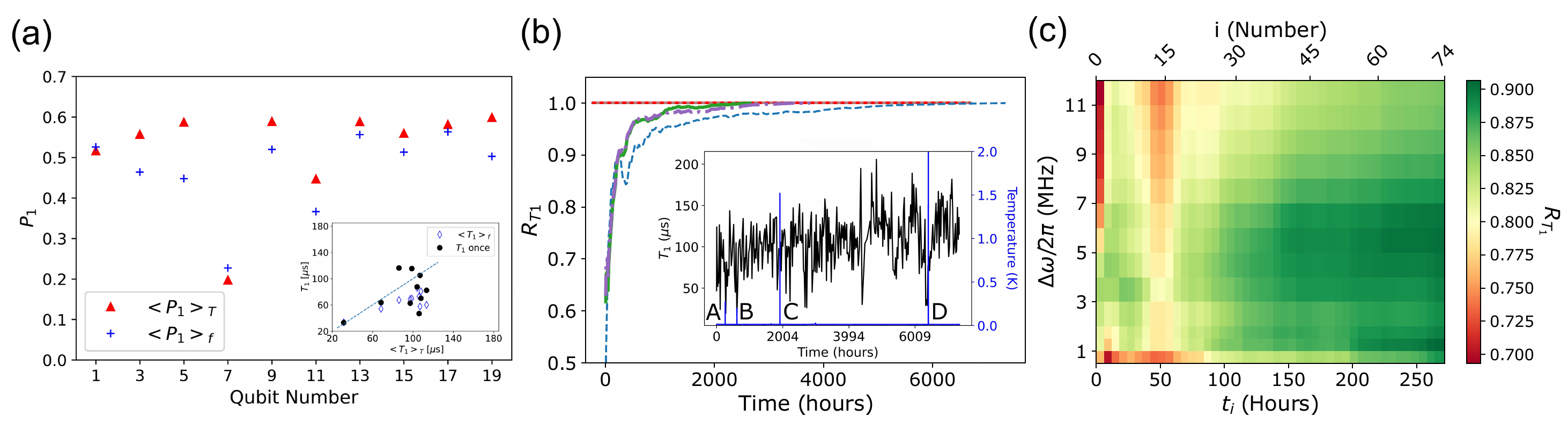}
    \caption{(a) Comparison of $\langle P_1 \rangle_{\omega,t}$ at $50 \mu$s and $\langle P_1 \rangle_T$ averaged for $\sim$9 months and evaluated at a $\tau$ of 53 $\mu$s decay time. $\langle P_1\rangle_{\omega,t}$, is averaged over $\Delta \omega/2\pi$ = 5 MHz after a single measurement that took $\sim$20 minutes. (inset) A scatter plot using $\langle T_1 \rangle_T$'s averaged over 9 months of measurement as the dependent variable and $\langle T_1 \rangle_{\omega,t}$ or $T_1$s from a single day. The line is a guide to the eye showing a 1:1 correlation. (b) The Pearson $R$ dependence on time averaging of the $T_1$'s of the odd numbered qubits up to time, T, {\color{black} for three cases: (i) the entire time series (dash), (ii) the time series between temperature excursion B and D (dash-dot), lettered locations indicated in the inset, and (iii) the time series between C and D for which no temperature excursions were recorded (solid). The intermediate time series are shifted in time index to compare more directly at short times with the full time series. The differences in $R$ are within the standard deviation calculated for sampling $T_1$ time series with a Gaussian distributed range of values, see appendix \ref{PearsonCorVstd}. (inset) The $T_1$ (black, left) and mixing chamber temperature (blue, right) time series for $Q_{13}$. Spacing of measurements is non-uniform. The minimum spacing is $\sim$24 hours apart. Each temperature excursion is labeled with a letter.} (c) Pearson correlation, $R$, dependence on time slice averaging and frequency range, $\Delta \omega$, of the odd numbered qubits.}
    \label{fig:Fig4}
\end{figure*}

We compare $\langle P_1 \rangle_{\omega,t}$ to $\langle P_1 \rangle_T$ from the daily $T_1$ measurements over $T_{\mathrm{max}} \sim$ 9 months evaluated at $\tau = 53 ~\mu s$, shown in Fig. \ref{fig:Figure1}. The $\langle P_1 \rangle_{\omega,t}$ are calculated for a $T_1$ delay time of $\tau$ = 50 $\mu$s for 10 qubits in the device for the first time slice and a cutoff frequency $\Delta \omega/2\pi$ = 5 MHz. A qualitatively close agreement for all 10 qubits is observed, see Fig. \ref{fig:Fig4} (a).

\begin{table}[h!]
\begin{tabular}{||c |c ||} 
\hline
\makecell{Symbol}&\makecell{Definition} \\ 
\hline\hline
$\omega_q$ & Qubit frequency \\
\hline
{\color{black}$\Omega_{s,j}$} & {\color{black}$j^{th}$ Stark drive amplitude point }  \\
 & {\color{black}in a frequency scan (see eqn. \ref{StarkShift})} \\
\hline
$\omega_j$ & {\color{black} $j^{th}$ Qubit Stark shift,}  \\
 & {\color{black} $\Delta \omega_q(\Omega_{s,j})$ (see eqn. \ref{StarkShift}) }\\
\hline
$d\omega_j$ & Frequency bin size at the \\
 & Stark shifted frequency location \\
\hline
$\Delta \omega$ & Maximum qubit Stark shift \\
\hline
$N$ & Number of $T_1$ measurements \\
 & for $\sim$9 month time series {\color{black}in an average}\\
\hline
$n$ & {\color{black}Total number of spectroscopy} \\
 & {\color{black}time slices in a moving average}\\
\hline
$T_i$ & {\color{black}Time of $i^{th}$ time bin} for the  \\
 & $\sim$9 month $T_1$ time series \\
\hline
$t_i$ & {\color{black} Time of $i^{th}$ time bin } \\
 & of the spectroscopy time series  \\
\hline
{\color{black}$T_1(T_i)$} & {\color{black} $T_1$ measured at $i^{th}$ time} \\
\hline
$\tau$ & Decay time at \\
 & which $P_1$ was evaluated \\
\hline
$P_1(\omega_q+\omega_j,\tau,t_i)$ & Probability of $|1\rangle$ at $\tau$ {\color{black}delay} for\\
 & time slice $t_i$ and frequency shift $\omega_j$ \\
\hline
$\langle P_1 \rangle_{T}$ &  Probability of $|1\rangle$ at $\tau$\\
 & averaged over $\sim$9 month time series \\
\hline
$\langle P_1 \rangle_{\omega,t}$ &  Probability of $|1\rangle$ at $\tau$  averaged over \\
 & frequency and spectroscopy time \\
\hline 
{\color{black}
$\langle T_1 \rangle_{T_{0 \rightarrow N}}$} & {\color{black}Moving average of $T_i$} \\ 
 & {\color{black}measurements from $T_0$ to $T_n$ }\\
\hline
$\langle T_1 \rangle_{T}$ & $T_1$ averaged over entire $\sim$9 months \\
\hline
$\langle T_1 \rangle_{\omega,t}$ & $T_1$ average from $\langle P_1 \rangle_{\omega,t}$ \\
\hline
$Q_k$ & k'th qubit  \\
\hline
$R(t_i)$ & $\langle T_1 \rangle_T$ correlation to $\langle T_1 \rangle_{\omega}$ at $t_i$ for \{$Q_k$\}   \\
\hline
$R(T_i)$ & $\langle T_1 \rangle_T$ correlation to $T_1$ at a single time \\  
  & $T_i$ of the $\sim$9 month series for \{$Q_k$\}\\
\hline
{\color{black}$\langle R \rangle_{t_{0 \rightarrow n}}$} & {\color{black} $\langle T_1 \rangle_T$ correlation to $\langle T_1 \rangle_{\omega,t_{0 \rightarrow n}}$ for \{$Q_k$\}}\\
\hline
{\color{black}$\langle R \rangle_{T_{0 \rightarrow N}}$} & {\color{black} $\langle T_1 \rangle_T$ correlation to $\langle T_1 \rangle_{T_{0 \rightarrow N}}$ for \{$Q_k$\}} \\
\hline

\end{tabular}

\caption{List of symbols}
\label{table:paperdefs}
\end{table}

A $\langle T_1 \rangle_{\omega,t}$ can also be estimated for each $\langle P_1 \rangle_{\omega,t}$ at $\tau$  = 50 $\mu$s by assuming an exponential decay. The approximate equivalence of $\langle T_1 \rangle_{\omega,t}$ and $\langle T_1 \rangle_T$ is seen in the scatter plot of Fig \ref{fig:Fig4} (a) inset. {\color{black}A near 1:1 relationship is observed when this approach is applied more broadly across many IBM devices, see Appendix \ref{OneToOne}.} Furthermore, the poorer correlation between $\langle T_1 \rangle_T$ and a single instance of $T_1$ measurements, is also shown by larger scatter, as seen in Fig \ref{fig:Fig4} (a) inset.

To quantify with a single value {\color{black}the correlation between $\langle T_1 \rangle_T$ or $\langle P_1 \rangle_T$ and their estimators for many qubits, we use a Pearson $R$ measure across the ten odd-labeled qubits, 
\begin{equation}
   R = 
   \frac{\sum\limits_{k=0}^{d-1}(\langle X\rangle_{T,Q_i}-\overline{\langle X\rangle_{T}})(\langle X\rangle_{\omega,t,Q_i}-\overline{\langle X\rangle}_
   {\omega,t})}{\sqrt{\sum\limits_{k=0}^{d-1}(\langle X\rangle_{T,Q_k}-\overline{\langle X\rangle_{T}})^2\sum\limits_{k=0}^{d-1}(\langle X\rangle_{\omega,t,Q_i}-\overline{\langle X\rangle}_{\omega,t})^2}}
\end{equation}
where d is the number of qubits in the device or analysis, 10 in this case, and X is the observable $P_1$ or $T_1$. The Pearson correlation is a normalized covariance between two variables reflecting a linear correlation from 1 to -1, where $R$ = 1 (-1) represents a $100\%$ positive (negative) correlation and $R$ = 0 indicates no correlation. Strong R correlation can therefore signal the existence of a potential linear mapping between the estimator and $\langle T_1 \rangle_T$, in particular, possibly one that is 1:1 or a scaling factor that will reliably estimate $\langle T_1 \rangle_T$}. 

For a single frequency sweep that takes $\sim$20 minutes, we obtain 0.76 $<$ {\color{black}$R(t_i)$} $<$ 0.84 correlation between $\langle T_1\rangle_{T}$ and $\langle T_1\rangle_{\omega,t}$ for 0.5 MHz $<$ $\Delta \omega$ $<$ 5 MHz. Using the $P_1$ values without assuming an exponential dependence leads to stronger correlations of 0.87 $<$ {\color{black}$R(t_i)$ }$<$ 0.91. Both of these are substantially stronger than the correlation found between the representative instance of $T_1$ and $\langle T_1\rangle_{T}$, which was $R$ = 0.29. We note this instance of $R$ can have a large spread, as seen by simulations {\color{black}of Gaussian distributed fluctuations} in Appendix \ref{PearsonCorVstd}.

{\color{black}A better estimate of the $\langle T_{1} \rangle_T$ for each qubit, $Q_k$, in the device can be obtained from a moving average of multiple, $N$, measurements. We show the evolution of $\langle R\rangle_{T_{0 \rightarrow N}} $ using a moving average of the $T_1(T_i)$ measurements, $\langle T_1 \rangle_{T_{0 \rightarrow N}}$, for each qubit, Fig. \ref{fig:Fig4} (b). The $\langle R\rangle_{T_{0 \rightarrow N}}$ exceeds $R$ $\sim$ 0.8 (i.e., strong correlation) after $\sim$10 measurements}, corresponding to a time exceeding 100 hours. {\color{black} Approximately 10 independent measurements is sufficient for fluctuations with magnitude $\sim$0.2$\langle T_1 \rangle_T$ to obtain a strong correlation, R $\sim$ 0.8, between an estimator (e.g., $\langle T_1 \rangle_{T_0 \rightarrow N}$) and $\langle T_1\rangle_T$. The details of R dependence on fluctuation magnitude and number of measurements in the moving average are discussed more completely in Appendix \ref{PearsonCorVstd}.}

{\color{black}Autocorrelation between $T_1(T_i)$ and $T_1(T_{i-1})$ measurements is an underlying challenge to fast estimation of $\langle T_1 \rangle_T$}. Evidence of autocorrelation can be seen for example in long term drifts in the average and short term correlations between $T_1$, inset of Fig. \ref{fig:Fig4} (b). On shorter time scales, our experimental data shows evidence of stronger {\color{black}autocorrelation} frustrating faster accurate estimation of $\langle T_1 \rangle_T$ and that the fastest $R \sim$ 0.8 can be obtained on order of 1-2 days, see appendices \ref{autocorrelation} and \ref{Rspec}. {\color{black}We conclude that $\langle T_1 \rangle_{\omega,t}$ shows promise as a method for faster estimation of $\langle T_1 \rangle_T$ than repeated  $T_1(\omega = \omega_q)$ measurements at only the qubit frequency. Extending the $\langle T_1 \rangle_{\omega,t}$ estimator to a set of many qubits, \{$Q_k$\}, in a device result in larger R, in the same time, compared to relying only on $T_1(\omega_q)$ measurements for each qubit. The R value simply being a  quantitative single value expression of the high correlation between each $\langle T_1 \rangle_{\omega,t}$ and $\langle T_1 \rangle_T$ across the entire set of qubits.} 

It is important to note that our calculations of $\langle T_1\rangle_{\omega,t}$ employ an equal weighting of $P_1$ associated with every frequency bin {\color{black} and the same choice of $\Delta \omega$ for every qubit}. However, it is not \textit{a priori} clear that equal weighting is a representative choice over the $\Delta \omega$ range. For example, how evenly does the spectral diffusion of each TLS contribute to the $T_1$ of the qubit? The strong correlation of $\langle T_1\rangle_{\omega,t}$ with $\langle T_1\rangle_{T}$ with equal weighting suggests that an ergodic-like sampling of the TLSs near the qubit frequency is a reasonable first approximation. {\color{black} The ergodic behavior of the $T_1$ estimators is examined more completely in Appendix \ref{T1ergodicity} and Appendix \ref{T1fcErgodicity}. Central to the question of assigning a $T_1$ estimate to any qubit, we observe that $\langle T_1 \rangle_T$ behaves ergodically for all the qubits despite short term $1/f^{\alpha}$ correlated behavior (i.e., a constant mean $\langle T_1 \rangle_T$ can be identified). Assignment of any $T_1$ estimate could alternatively be made impossible in the presence of drift, which is not observed in these qubits, see Appendix \ref{TimeseriesStationarity} and Appendix \ref{T1ergodicity} for further details about weak stationarity and ergodicity}. Furthermore, the strong correlation of $\langle T_1 \rangle_T$ to $\langle T_1 \rangle_{\omega,t}$ using only the $P_1(\omega,\tau,t)$ spectrum around the qubit is consistent with a leading hypothesis that the $\langle T_1 \rangle_T$ is dominated by TLS behavior rather than other stochastic or static contributions.

 \section{Correlation dependence on frequency and measurement time}
 \label{section:Correlation dependence}
 {\color{black}A natural question about the estimator $\langle T_1\rangle_{\omega,t}$ is, what are the optimal parameter choices for frequency range $\Delta \omega$, $n$ autocorrelated samples and the spacing in time, $\Delta t = t_i - t_{i-1}$, to obtain sufficiently weakly autocorrelated measurements and a fast, accurate measure of $\langle T_1 \rangle_T$.} Since the optimum choices are presently not known \textit{a priori}, we evaluate and plot {\color{black} $\langle R \rangle_{t_{0 \rightarrow n}}$} versus $\Delta \omega$ and $t_i$ in Fig. \ref{fig:Fig4} (c) {\color{black}to guide future application of this approach.} Equal frequency bin weighting of $P_1$ {\color{black}is used. While this order of magnitude choice of $\Delta \omega$} produces a reasonably good first approximation for correlation across the entire range, the plot displays several unexplained features (e.g., non-monotonic dependence on $\Delta \omega$) indicating the unsurprising insufficiency of these two {\color{black}globally applied} parameters (i.e., $\Delta \omega$ and $t$) alone to weight the frequency contribution of all the qubits and approach $R \sim$ 1. 
 {\color{black} Additional sensitivity analysis in Appendix \ref{T1fcErgodicity} also examines correlation between frequencies and highlights that individual qubits have different sensitivity to the range sampled, $\Delta \omega$. We see that a wide span of $\Delta \omega$ produces high $\langle R \rangle_{t_{0 \rightarrow n}}$, comparable or better than $R(T_i)$ from a single $T_1(\omega_q)$ measurement. We further show that not only is there a strong R correlation (e.g., linear dependence) but that $\langle T_1 \rangle_{\omega,t}$ approaches 1:1 quantitative agreement with $\langle T_1 \rangle_T$. The degree to which a $T_1$ estimator, from sampling the nearby frequency space, is quasi-ergodic and would converge to 1:1 agreement is addressed in much more detail in Appendix \ref{T1fcErgodicity} and Appendix \ref{OneToOne}.}
 
\section{Discussion: implications for process characterization}
\label{process}

The strong correlation between $\langle T_1 \rangle_{\omega,t}$ and $\langle T_1 \rangle_{T}$ suggests that long time $T_1$ averages might be estimated relatively rapidly using spectroscopy. This is in contrast to overcoming correlation times in $T_1$ at a single $\omega_q$ to obtain a representative $\langle T_1 \rangle_T$ for the qubit. 

{\color{black} Identification of better choices of $\Delta \omega$ and $n$ in this study were made with pre-knowledge of what $\langle T_1 \rangle_T$ was. These parameters will have to be chosen without this pre-characterization for future implementation of this method. Encouragingly, the R dependence on both these parameters} appears to be relatively weak suggesting that a heuristic choice for a single $\Delta \omega$ and $n$ might be sufficient to obtain useful estimates (i.e., $R > 0.8$) of $\langle T_1 \rangle_{T}$ for new processes when using this simple equal weighting approach until improved choices can be formulated (i.e., different frequency spans for each qubit and or weighted averaging over frequency).

{\color{black} More specifically we observe that $\mathcal{O}$(10) independent measurements is sufficient to obtain an R $\sim$ 0.8 or higher, see Appendix \ref{PearsonCorVstd}. We conjecture that one can obtain 10 approximately independent samples, $S$, in a single scan by sampling at frequency spacings, $\chi$, that are greater than the autocorrelation frequency width (i.e., a frequency spacing where correlation drops below $\sim$0.2). In this work, we found the correlation to become weak for $\mathcal{O}$(1 MHz), see Appendix \ref{T1fcErgodicity}. Then by this heuristic, a single spectroscopy scan would require a $\Delta \omega$ = $\frac{(S-1)}{2}\chi$, where $S$ = 10 for the target of $R \sim$0.8. We assume one of the measurements is done at the qubit frequency, $T_1(\omega_q)$, so for a $\chi \sim$ 1 MHz, a scan from $\pm$4.5 MHz would be suggested by such a heuristic. Extra $n$ measurements can be obtained by waiting longer than the autocorrelation time. The autocorrelation width, furthermore, can be evaluated in the same scan as that used for the $\langle T_1 \rangle_T$ estimate as long as a sufficiently wide range is sampled. Alternatively a second scan can be taken if the initial $\Delta \omega$ guess was too small.

Empirically we see diminishing gains in using ever larger $\Delta \omega$. Further research is needed to guide better limits on $\Delta \omega$ beyond the operational observation that $S \sim \mathcal{O}(10)$ produces a quasi-ergodic result for qubits with $\langle T_1 \rangle_T$ in the range of 10-200 $\mu$s, see Appendix \ref{T1fcErgodicity} for more details on quasi-ergodicity. Since we do find $\sim$1:1 agreement using a relatively small $\Delta \omega \sim$ 10 MHz for the $\sim$9 month time series and we observe that the distribution of $T_1(\omega_q,T_i)$ produces a constant standard deviation, see Appendix \ref{TimeseriesStationarity},} rather than growing (e.g., proportional to a random walk $\propto \sqrt{t}$), we speculate that optimal $\Delta \omega$ is bounded rather than growing indefinitely from spectral diffusion processes. Notably Klauder et al. calculate that dipole coupled ensembles that are proposed for TLS spectral diffusion \cite{black_spectral_1977}, will produce a truncated linewidth \cite{Klauder_PhysRev.125.912}.

\section{Conclusion}
In this work, we probe the temporal and spectral dynamics of superconducting qubit relaxation times. We study these dynamics in high coherence, single-junction transmons by developing a technique for energy relaxation spectroscopy of defect TLSs via the AC Stark effect. Our technique requires no additional hardware resources and can be easily sped up further by integration with reset schemes. Autocorrelation of $T_1$ frustrates rapid characterization of the long-time average $\langle T_1 \rangle_T$ and therefore accurate characterization of devices. Our analysis of the dynamics identifies a strong correlation between $\langle T_1 \rangle_T$ and its short time average over the local frequency span, $\langle T_1 \rangle_{\omega,t}$. The strong correlation of $\langle T_1 \rangle_T$ with $\langle T_1 \rangle_{\omega,t}$ is also consistent with a TLS dominated $T_1$ that quasi-ergodically samples the qubit local frequency neighborhood in contrast to static or uncorrelated stochastic processes. This work opens up several new promising directions for rapid process characterization and evaluation of device stability.

\section{Data Availability}
The data that support the findings of this study are available from the corresponding author on reasonable request.

\section{References}

\bibliography{MyLibrary}

\section{Acknowledgements}
We acknowledge technical support on the ibmq\_almaden device from the IBM Quantum deployment team. Additional insightful discussions, suggestions and assistance came from Nick Bronn, Andrew Cross, Oliver Dial, Doug McClure, Easwar Magesan, Hasan Nayfeh, James Raferty, Martin Sandberg, Srikanth Srinivasan, Neereja Sundaresan, Jerry Tersoff, Ben Fearon, Karthik Balakrishnan, James Hannon and Jerry Chow. 

MC also acknowledges support from Princeton Plasma Physics Laboratory through the Department of Energy Laboratory Directed Research and Development program and contract number DE-AC02-09CH11466 to complete parts of the analysis and manuscript.

\section{Author Contributions}
S. R. and A. K. developed the technique with contributions from M. C., I. L. and P. J. M. C. and S. R. performed the experiments. M.C., S.R and A. K. analyzed the data. M. C., S. R. and A.K, wrote the manuscript with feedback from the other authors.


\section{Competing Interests}
The authors declare that elements of this work will be included {\color{black} in patents} filed by the International Business Machines Corporation with the US Patent and Trademark office. The authors declare no other financial or non/financial competing interests in relation ot this published work.

\appendix
\section{Spectroscopy of odd numbered qubits in device}
\label{other spectroscopy}
\begin{figure*}[bh]
	\includegraphics[width=\textwidth, height=10cm, clip,trim = 0.0mm 0.0mm 0.0mm 0.0mm]{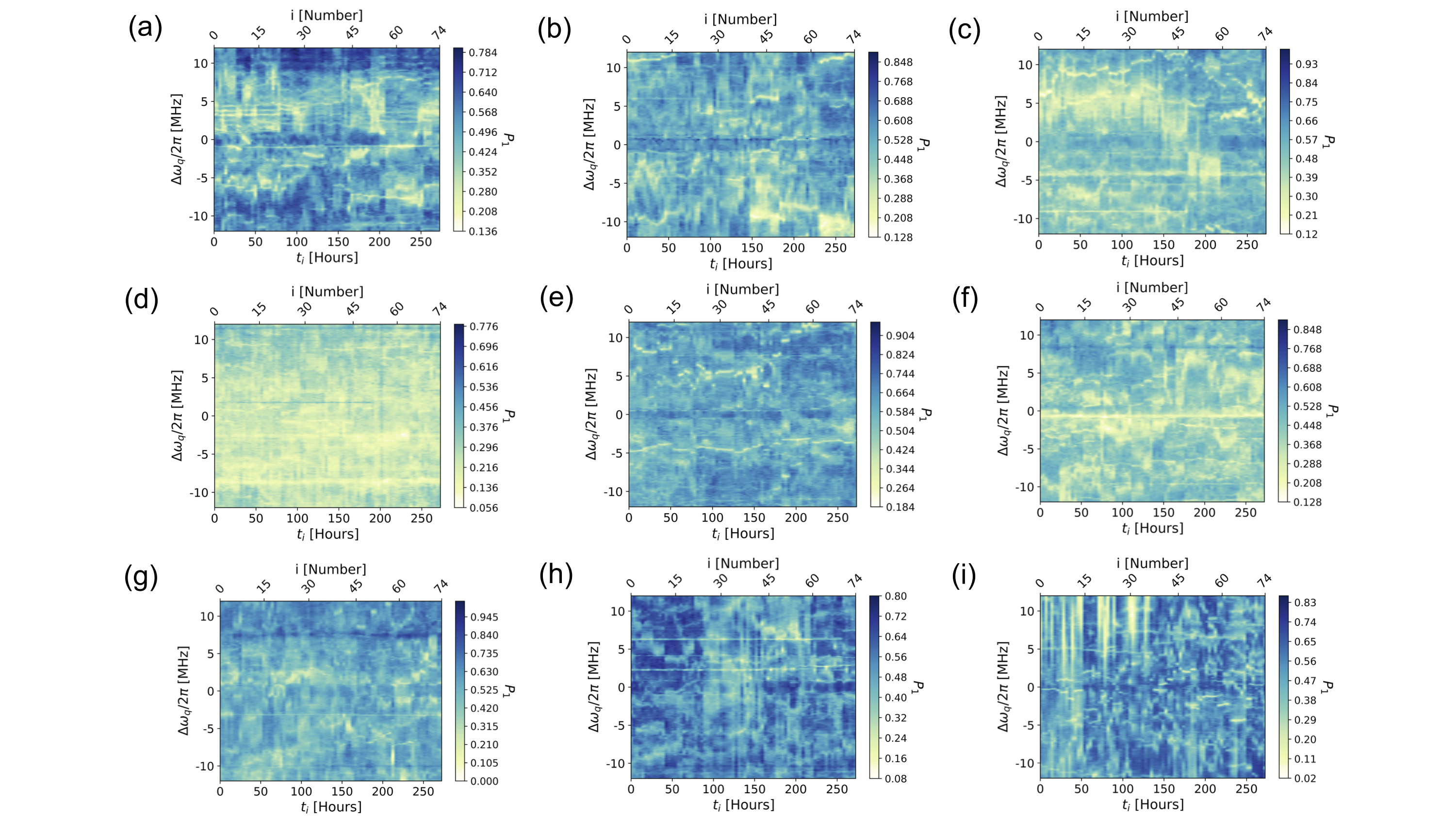}
	\caption{(a)-(i) Spectroscopy for odd qubits 1-19 excluding qubit 15 {\color{black} in respective order. Explicitly: (a) $Q_1$, (b) $Q_3$, (c) $Q_5$, (d) $Q_7$, (e) $Q_9$, (f) $Q_{11}$, (g) $Q_{13}$, (h) $Q_{17}$, (i) $Q_{19}$. $Q_{15}$} is shown in the main text.}
	\label{fig:FigE1}
\end{figure*}

The spectrally and temporally resolved dynamics of $T_1$ for all the odd numbered qubits in the device are provided for reference in Fig. \ref{fig:FigE1}.
The data was taken under the same conditions and at the same time as Fig. \ref{fig:Figure4} shown in the main body of the paper. Care was taken to avoid frequency collisions both over the range of qubit frequency shift and the placement of $\omega_s$. We note that the odd qubits do not have direct connectivity with each other, making cross talk effects negligible for these measurements.

\section{Stationarity of $T_{1}$ time series}
\label{TimeseriesStationarity}
{\color{black} 
An important question for $T_1$ analysis of qubits is whether there is in fact a representative $T_1$ mean and variance that can be assigned to each qubit. In time series analysis, this concept is described as weak stationarity \cite{Krishnan2015PnRProcess}. Weak stationarity is also a necessary condition for ergodicity \cite{MilottiReview}. 

The mean and standard deviations qualitatively appear relatively constant. A histogram of the $T_1$s over 9 months for $Q_{13}$ is plotted for a representative case along with a fit to a normal distribution Fig. \ref{fig:GaussDistT1s} (a). Visually, the distributions appear relatively normal indicating that the mean is not drifting substantially relative to the variance, while the skew and kurtosis are also relatively small. In other qubits, similar near normally distributed $T_1$ fluctuations are observed, see table \ref{table:MomentsTable}. The skew and kurtosis are frequently not discernible statistically from a normal distribution \cite{Kurtosistest_10.1093/biomet/70.1.227,Skewtest_doi:10.1080/00031305.1990.10475751} with notable exceptions such as $Q_7$, which has a tight distribution around the mean and some instances for which the skew was distinguishably larger than normal in $Q_3$ and $Q_{17}$.

\begin{table}[bh]
\begin{tabular}{||c |c |c |c |c | c |c ||} 

\hline
\makecell{Qubit} & \makecell{$\mu_{(1)}$ \\ ($\mu$s)} &\makecell{$\mu_{(2)}$ \\ ($\mu$s)} &\makecell{$\mu_{(3)}$} &\makecell{$\mu_{(4)}$} & \makecell{p-value \\ $\mu_{(3)}$ test} & \makecell{p-value \\ $\mu_{(4)}$ test}\\
\hline\hline
1 & 86.1 & 28.3 & -0.12 & -0.32 & 0.37 & 0.23\\
\hline
3 & 97.2 & 26.4 & -0.44 & 0.30 & 0.002 & 0.25\\
\hline
5 & 107.0 & 28.5 & -0.25 & 0.23 & 0.07 & 0.32\\
\hline
7 & 31.4 & 5.6 & -0.83 & 1.92 & $\sim$0 & $\sim$0\\
\hline
9 & 107.9 & 29.3 & -0.18 & -0.03 & 0.20 & 0.94\\
\hline
11 & 68.4 & 17.7 & -0.11 & -0.09 & 0.42 & 0.88\\
\hline
13 & 106 & 32.7 & -0.09 & 0.22 & 0.50 & 0.34\\
\hline
15 & 98.8 & 28.4 & -0.19 & -0.32 & 0.17 & 0.23\\
\hline
17 & 104.1 & 28.3 & -0.28 & -0.32 & 0.04 & 0.22\\
\hline
19 & 113.4 & 35.1 & -0.14 & -0.24 & 0.29 & 0.44\\
\hline

\end{tabular} 
\caption{{\color{black} Calculated first four moments of the $T_1$ distributions for all the odd qubits. The mean, standard deviation, skew and kurtosis are indicated as $\mu_{(1)}$, $\mu_{(2)}$, $\mu_{(3)}$, and $\mu_{(4)}$, respectively, in the table. The table also contains the p-value results of skew and kurtosis tests for normality \cite{Kurtosistest_10.1093/biomet/70.1.227,Skewtest_doi:10.1080/00031305.1990.10475751}. The null hypothesis is that the test distribution comes from a normal distribution. A value $\geq$ 0.05 is often used as the threshold for the null hypothesis to be accepted (i.e., the distribution is normal-like for that tested moment).}}
\label{table:MomentsTable}
\end{table}

We also show the moving average of the standard deviation of the $T_1$ distributions for each qubit in Fig. \ref{fig:GaussDistT1s} (b). The standard deviation of each qubit is normalized to its respective mean, $\mu$. We see that the general trend is for the standard deviations to converge towards their mean, $\sigma \sim \mathcal{O}$($\mu$). Drift in $\sigma$ is small relative to $\mu$. The $\sigma$ behaves weakly stationary rather than, for example, random walk like (i.e., $\sigma \propto \sqrt{t}$).   

\begin{figure}[ht]
	\includegraphics[width=85mm, clip,trim = 0.0mm 60.0mm 0.0mm 40.0mm]{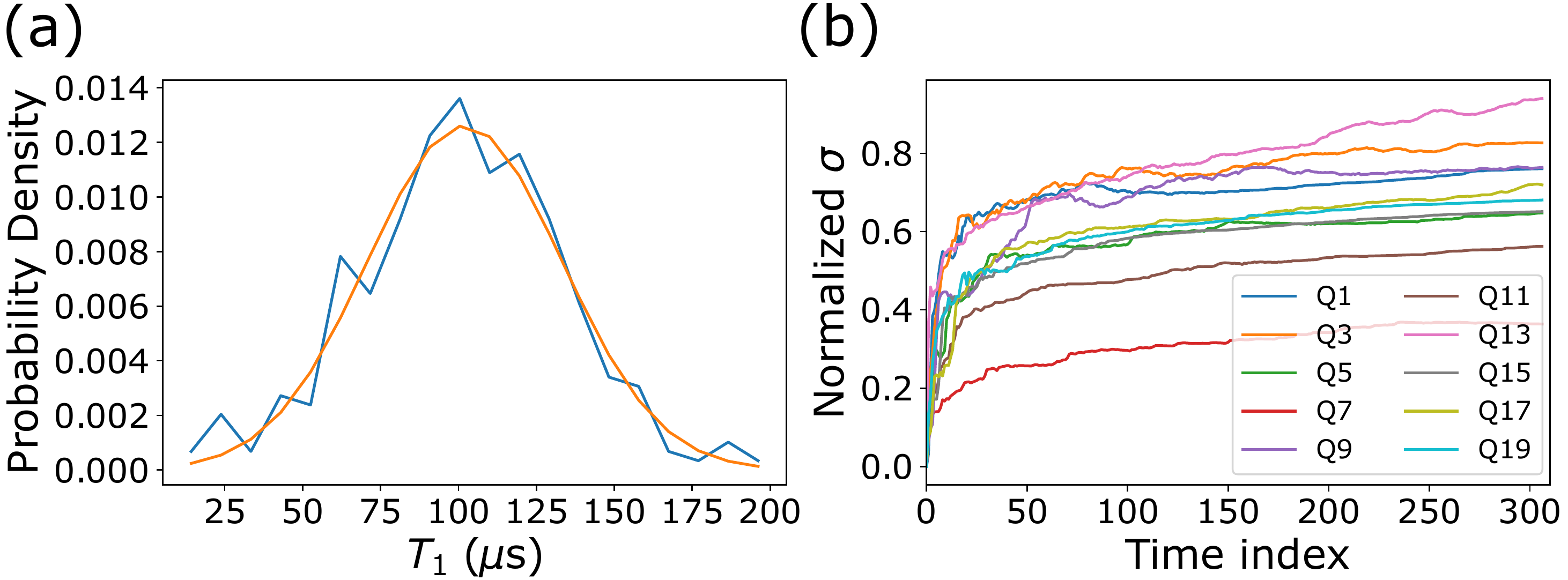}
	\caption{{\color{black} (a) Probability density of $T_1$ fluctuations for a representative qubit, $Q_{13}$. The normalized histograms are derived from the $T_1$ time series. (b) Normalized cumulative moving average of the standard deviation of the $T_1$ fluctuations as a function of time index spanning $\sim$9 months.}}
	\label{fig:GaussDistT1s}
\end{figure}

To more rigorously test the weak stationarity, we apply an augmented Dickey Fuller (ADF) test \cite{mackinnon_approximate_1994,MacKinnonAFD,HARRIS1992381} to the timeseries. The ADF test is a commonly used test of weak stationarity, testing both drift and constant variance. In our case we are most concerned with how to test whether the variance is stationary after observing that the mean is stationary, at least within $\mathcal{O}(\sigma_{T1})$ over 9 months. Random walks are the canonical case for which the mean is constant but the timeseries will be non-stationary because the variance grows with time. In particular, the ADF tests the likelihood of a unit root difference equation regression with the timeseries in question. Unit root is synonomous with random walk behavior. ADF uses the following parameterized model: 

\begin{equation}
    y_i = \alpha y_{i-1} + \sum_{j=1}^{\rho}\beta \Delta y_{i-j} + \delta + \gamma i + \epsilon_i 
\end{equation}
where i is the time step index, $\alpha$ is the root, the sum is over additional lag terms, $\delta$ is a constant offset, $\gamma$ is the slope of a linear trend and $\epsilon$ is a random error term that is normally distributed with a standard deviation of $\sigma_{\epsilon}$. The lag terms importantly account for effects of serial correlation (e.g., non-Markovian behavior expected in 1/f noise), while the drift term can be used to establish 'trend stationary' behavior. The null hypothesis, $H_0$, is that there is a unit root. If the time series is stationary, the ADF $H_0$ must be rejected. 

We test using $\gamma$ = 0, which will accept the null hypothesis, $H_0$, for either the case of a unit root (e.g., random walk) or for an $\mathcal{O}(\sigma_{\epsilon})$ non-stationary trend summed with a random error term. Table \ref{table:ADFtable} shows the results of the tests. All $T_1$ timeseries reject the ADF $H_0$. All $T_1$ timeseries are therefore consistent with being weakly stationary.  

\begin{table}[b!]
\begin{tabular}{||c |c |c ||} 

\hline
\makecell{Qubit} & \makecell{t-stat} &\makecell{p-value} \\
\hline\hline
1 & -9.3 & $9.1\times 10^{-14}$ \\
\hline
3 & -6.14 & $1.0\times 10^{-6}$   \\
\hline
5 & -10.1 & $1.6\times 10^{-15}$ \\ 
\hline
7 & -11.3 & $5.2\times 10^{-18}$ \\
\hline
9 & -12.1 & $2.5\times 10^{-19}$ \\
\hline
11 & -8.5 & $5.8\times 10^{-12}$ \\
\hline
13 & -6.0 & $1.9\times 10^{-6}$  \\
\hline
15 & -9.9 & $3.0\times 10^{-15}$  \\
\hline
17 & -9.6 & $1.4\times 10^{-14}$  \\
\hline
19 & -4.1 & $7.3\times 10^{-3}$  \\
\hline

\end{tabular} 
\caption{{\color{black}ADF tests for odd qubit $T_1$ time series using the python statsmodel library. The tests use a Bayesian information criteria to determine the number of lags. P-values of greater than 0.05 are the typical statistical threshold for acceptance of the $H_0$, null hypothesis. The table values are for $\gamma$ = 0.}}
\label{table:ADFtable}
\end{table}

} 

\section{Pearson correlation dependence on standard deviation and sampling}
\label{PearsonCorVstd}


We calculate an expectation of how many uncorrelated measurements are necessary to achieve strong $R$ correlation (i.e., $R \sim$ 0.8). We calculate a Pearson {\color{black}$\langle R(n_q,\sigma_{mk}) \rangle_{T_{0 \rightarrow N}}$ for $n_q$ being a ten qubit device, $\sigma_{mk}$ being the $T_1$ standard deviation for the $k^{th}$ }qubit and where $N$ is the number of uncorrelated $T_1$ measurements for each of the ten qubits. The simulated {\color{black}$\langle R \rangle_{T_{0 \rightarrow N}}$} calculation has a long-time average $T_1$ assigned to each of the 10 qubits, $\langle T_1 \rangle_{T,k}$. For every $k$-th qubit, $\langle T_1 \rangle_{T,k}$ is chosen randomly from a normal distribution with a mean of 100 $\mu$s and a standard deviation of 10 $\mu$s. $\langle T_1 \rangle_{T,k}$ simulates a long time stationary $\langle T_1 \rangle_T$ for each qubit. This standard deviation is representative of process variation in the qubits of each simulated device. 

We then simulate a sequence of measurements. Each measurement obtains an instantaneous $T_1$ for each qubit in the device. The $T_1$ measurement is chosen from a normal distribution with a {\color{black} standard deviation of $\sigma_{mk}$ = 0.2$\langle T_1 \rangle_{T,k}$, which is of similar magnitude to what is observed in the device. The $\sigma_{mk}$ is a measure of the time fluctuating $T_1$ centered around $\langle T_1 \rangle_{T,k}$, in contrast to the 10 $\mu$s above, which is the the variability of the static $\langle T_1 \rangle_{T,k}$ centered around $\langle T_1 \rangle_{T,n_q}$ = $\frac{1}{n_q} \sum_k^{n_q} \langle T_{1} \rangle_{T,k}$.} 

We can obtain an estimate of the Pearson $R$ correlation between $\langle T_1 \rangle_{T,k}$ and a single measurement instance $T_{1,k}$ for all the qubits. The effect of multiple uncorrelated device measurements is then simulated by repeating the device measurement and updating the average of the $T_{1,k}$ with all previous $T_1$ measurements.  
In order to simulate the dependence of {\color{black}$\langle R \rangle_{T_{0 \rightarrow N}}$} on the number of uncorrelated measurements $N$, we first define an expectation value, { \color{black}$\langle \langle R \rangle_{T_{0 \rightarrow N}} \rangle_{Ndev}$}, averaged over {\color{black} $N_{dev}$ = 200 devices with 4, 10, 20 and 50 qubits.}

\begin{figure}
	\includegraphics[width=80mm, clip,trim = 0.0mm 0.0mm 0.0mm 0.0mm]{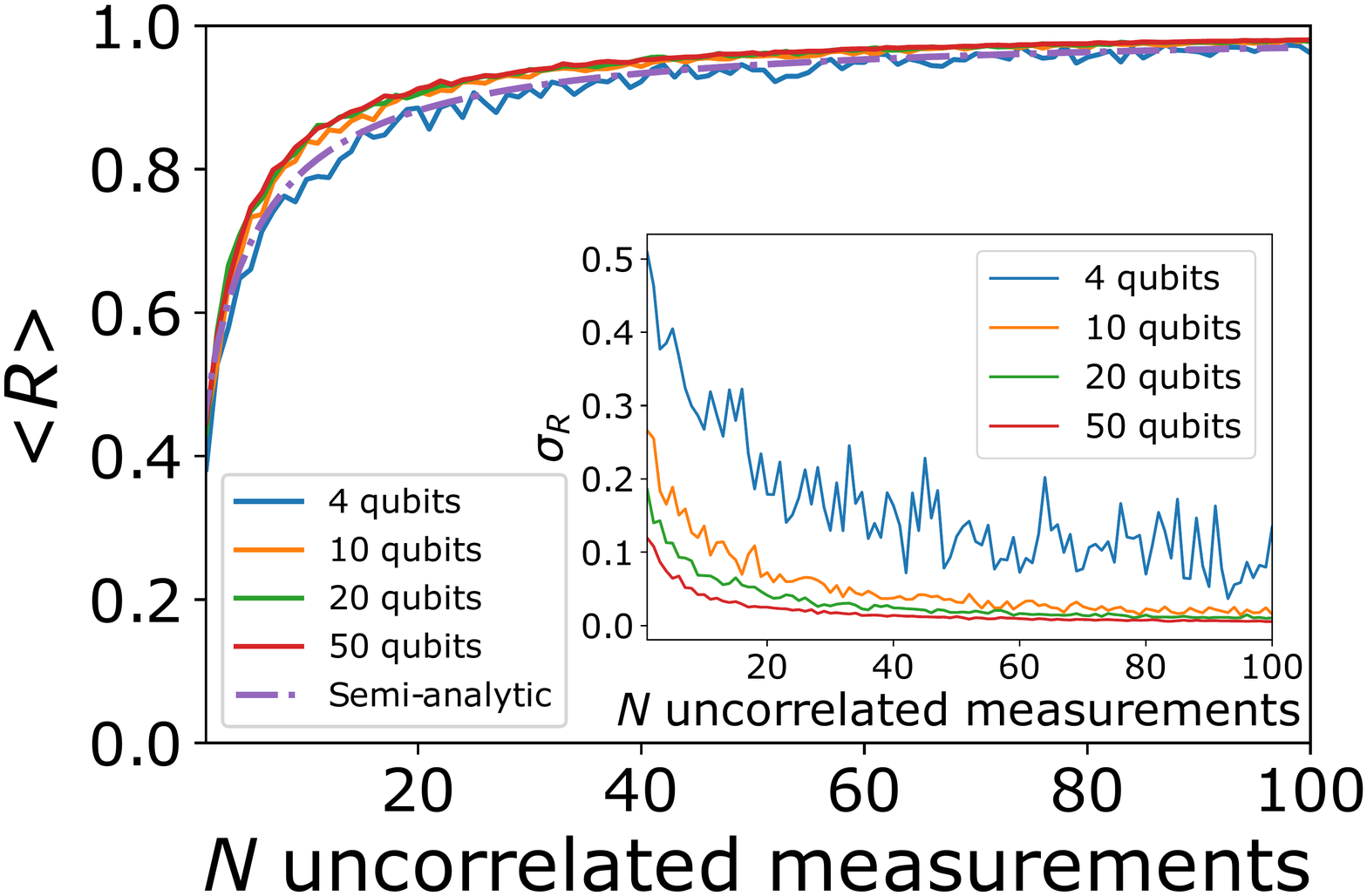}
	\caption{Simulated Pearson $R$ dependence {\color{black} on number of qubits in the device and} on $N$ uncorrelated device measurements of $T_1$ for each qubit in the device. The simulation is done for 200 simulated devices to find an expectation value $\langle R \rangle$. {\color{black}The semi-analytic expression in the text agrees well to the 4 qubit simulation.} The inset shows the dependence of the standard deviation of $R$ on $N$ {\color{black}and qubit number.}}
	\label{fig:FigRdep}
\end{figure}

{\color{black}$\langle \langle R \rangle_{T_{0 \rightarrow N}} \rangle_{Ndev}$} approaches unity with increasing uncorrelated measurements, as seen in Fig. \ref{fig:FigRdep}. This indicates that the averaging of uncorrelated measurements increasingly produces an accurate estimate of $\langle T_1 \rangle_{T,k}$. The standard deviation, $\sigma_R$, of {\color{black}$\langle \langle R \rangle_{T_{0 \rightarrow N}} \rangle_{Ndev}$} is also shown in the inset, Fig. \ref{fig:FigRdep}. We can see that initial {\color{black}$\langle R \rangle_{T_{0 \rightarrow N}}$} values can be very low, and around $N$ = 10 uncorrelated measurements are required to obtain $R \sim$ 0.8, the correlation obtained from the frequency averaging of $T_1$ discussed in the main text. 

{\color{black} To provide additional insight into the {\color{black}$\langle R \rangle_{T_{0 \rightarrow N}}$} dependence on $n_q$, $\sigma_{mk}$ and $N$ measurements, we derive an analytic expression for {\color{black}$\langle R(n_q,\sigma_{mk}) \rangle_{T_{0 \rightarrow N}}$}. We express R in terms of differences, $\delta X$, from $ \langle T_1 \rangle_{T,n_q} $.
Explicitly writing out the first few terms of $\langle R(n_q,\sigma_{mk}) \rangle$'s sums for a multiqubit device:
\begin{equation}
    \langle R(n_q,\sigma_{mk}) \rangle = \frac{\delta X_1 (\delta X_1 + \sigma_{m1}) + \delta X_2 (\delta X_2 + \sigma_{m2}) + ...}{\sqrt{[\delta X_1^2 + \delta X_2^2 + ...][(\delta X_1 + \sigma_{m1})^2 + ...]}} 
    \label{RpreExpression}
\end{equation}
where we define $\delta X_k = \langle T_1 \rangle_{T,k} - \langle T_1 \rangle_{T,n_q}$ and $\sigma_{mk}$ is the standard deviation for measurements of $T_{1,k}$ for the $k^{th}$ qubit in the device. 

We parameterize the standard deviation of the $T_1$ measurement with a scaling constant $\alpha$ for the $k^{th}$ qubit as $\sigma_{mk} = \alpha_i \langle T_{1} \rangle_{T,k}$. Likewise, we assume that the $\langle T_1 \rangle_{T,k}$ in a device are normally distributed and may be parameterized with $\beta_i$ as $\sigma_{di} = \beta_i \langle T_1 \rangle_{n_q}$ (i.e., $\langle T_1 \rangle_{T,k} = \langle T_1 \rangle_{n_q} + \sigma_{dk}$).

We solve for $\langle R \rangle_{T_{0 \rightarrow N}}$ for a particular device instance with $\langle T_{1} \rangle_{T,k}$, in the simpler case that $\alpha_k$ = $\alpha$ (i.e., $\sigma_{mk}$ = $\sigma_m$), and with the device defined by $\vec\beta$ = $\{ \beta_1, \beta_2,... \}$. The $\vec\beta$ defines the $\langle T_1 \rangle_T$s of a device instance. Multiplying and reorganizing the first few terms:

\begin{equation}
    \langle R(n_q,\sigma_{m}) \rangle = \frac{(\delta X_1^{2} + \delta X_2^{2} + ...) + \sigma_{m}(\delta X_1 + \delta X_2 + ...)}{\sqrt{[\delta X_1^2 + \delta X_2^2 + ...][(\delta X_1 + \sigma_{m})^2 + ...]}} 
    \label{RpreExpression2}
\end{equation}
Substituting $\delta X_k + \sigma_{m} \rightarrow \delta X_k + \frac{\sigma}{\sqrt{N}}$ to express the moving average dependence of $T_1$ on N measurements, we obtain:

\begin{equation}
    \langle R(n_q,\sigma_{m},N) \rangle = \frac{(\sum_i^{n_q}\beta_i^2) + \frac{\alpha}{\sqrt{N}}(\sum_i^{n_q} \beta_i) }{\sqrt{(\sum_i^{n_q}\beta_i^2)(\sum_i^{n_q}(\beta_i + \frac{\alpha}{\sqrt{N}})^2)}}
    \label{RNexpression3}
\end{equation}

To compare to the simulations in Fig. \ref{fig:FigRdep}, we numerically sample many instances of eqn. \ref{RNexpression3}, $N_{dev}$ instances of $\vec\beta$. The $\vec\beta$ are assumed to have a normal distribution. The expression shows good agreement with the full numerical simulation described above, Fig. \ref{fig:FigRdep}. The expression \ref{RNexpression3} provides the quantitative dependence of how increasing $N$ reduces the uncertainty in R through reducing the uncertainty in each $T_{1,k}$ estimator. The uncertainty is reduced by the familiar $\sqrt{N}$ dependence resulting in $\lim_{N \rightarrow \infty} \langle R \rangle_{T_{0 \rightarrow N}} \rightarrow 1$. Increasing the number of qubits in the device also can be used to reduce $\sigma_R$ for a single device measurement, playing a similar role of averaging $\vec\beta$ instead of alternatively measuring many devices.   


}

\section{Autocorrelation of $T_1$($\omega_q,T_i)$}
\label{autocorrelation}
\begin{figure}[hb]
	\includegraphics[width=85mm, clip,trim = 5.0mm 20.0mm 5.0mm 0.0mm]{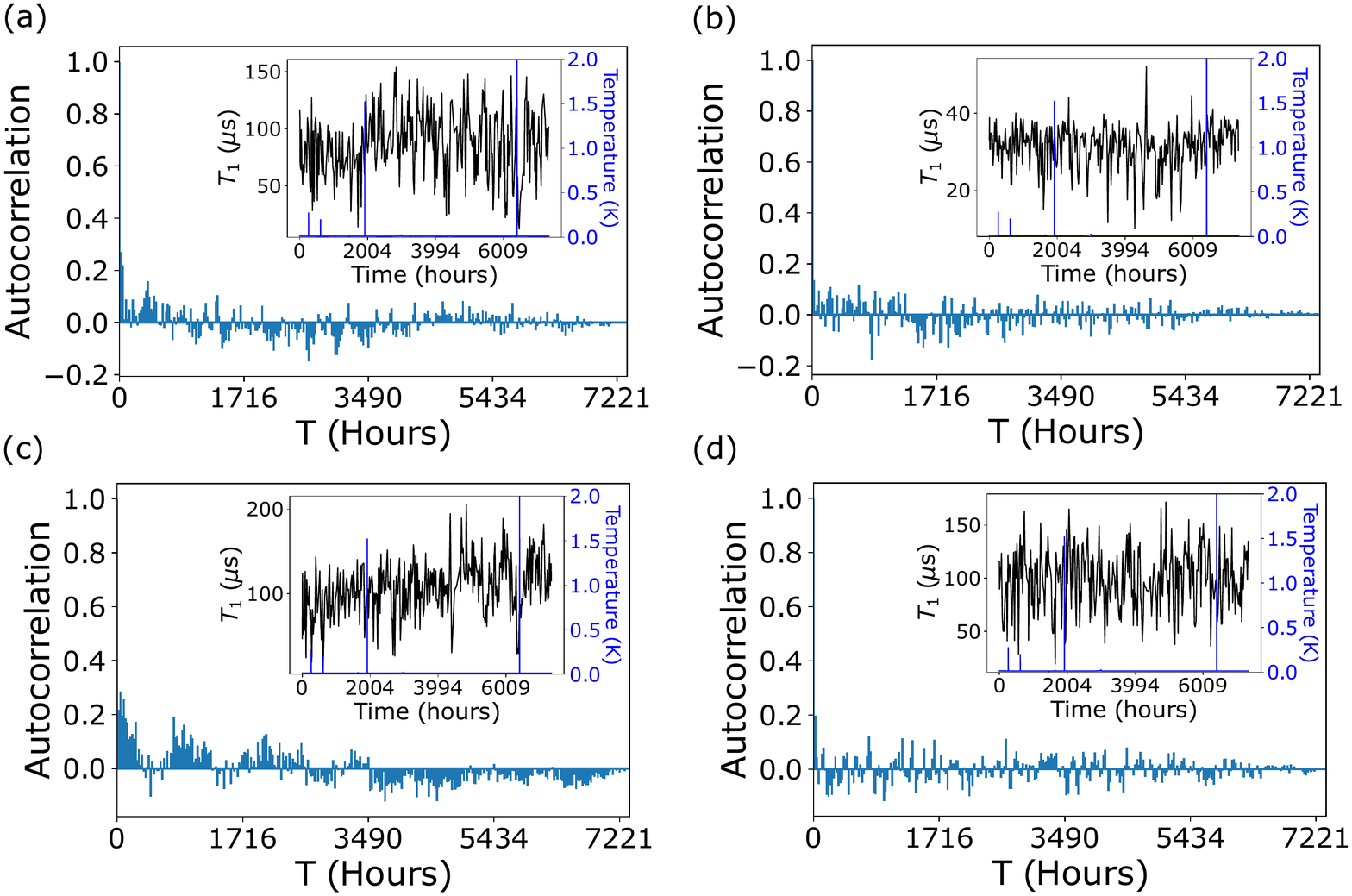}
	\caption{(a), (b), (c) and (d) Autocorrelation as a function of measurement lag time for qubits 1, 7, 13 and 15, respectively. The autocorrelation is detrended using the mean of the time series and normalized using the estimated variance. The insets show the $T_1$ time series from which the autocorrelation was calculated {\color{black}and the corresponding temperature of the mixing chamber plate.}}
	\label{fig:autocorrelation}
\end{figure}

We now ask the question: How long does it take to obtain an uncorrelated measurement of $T_1$? To address this, we calculate the autocorrelation for the $\sim$9 month $T_1$ time series, depicted in Fig \ref{fig:autocorrelation}.
The autocorrelation is detrended using the mean and normalized using the estimated variance. {\color{black} The period between sampling is approximately 24 hours with some variability (i.e., hours). All qubits have some time correlation in the first few measurements (i.e., 1-2 days). For short time behavior see appendix \ref{Rspec}. Some qubits also show weak and decaying autocorrelation at longer times. For example, $Q_{13}$ and $Q_1$'s time series appear to have long time trends, Fig. \ref{fig:autocorrelation}. The decaying oscillatory autocorrelation is qualitatively consistent with a mean reverting time series, for example, an Ornstein-Uhlenbeck process commonly used to simulate pink noise \cite{Singh_PhysRevE.98.012136}; and is also consistent with the weak stationarity of the same time series found in Appendix \ref{TimeseriesStationarity}.} In general, these longer timescale autocorrelations highlight a challenge in extrapolating $\langle T_1 \rangle_T$ from sampling $T_1$ at a single frequency at early times in the time evolution.

\section{Thermal excursion effects on autocorrelation}
\label{AutoTempExcursion}
\begin{figure}[hb]
	\includegraphics[width=85mm, clip,trim = 0.0mm 0.0mm 0.0mm 0.0mm]{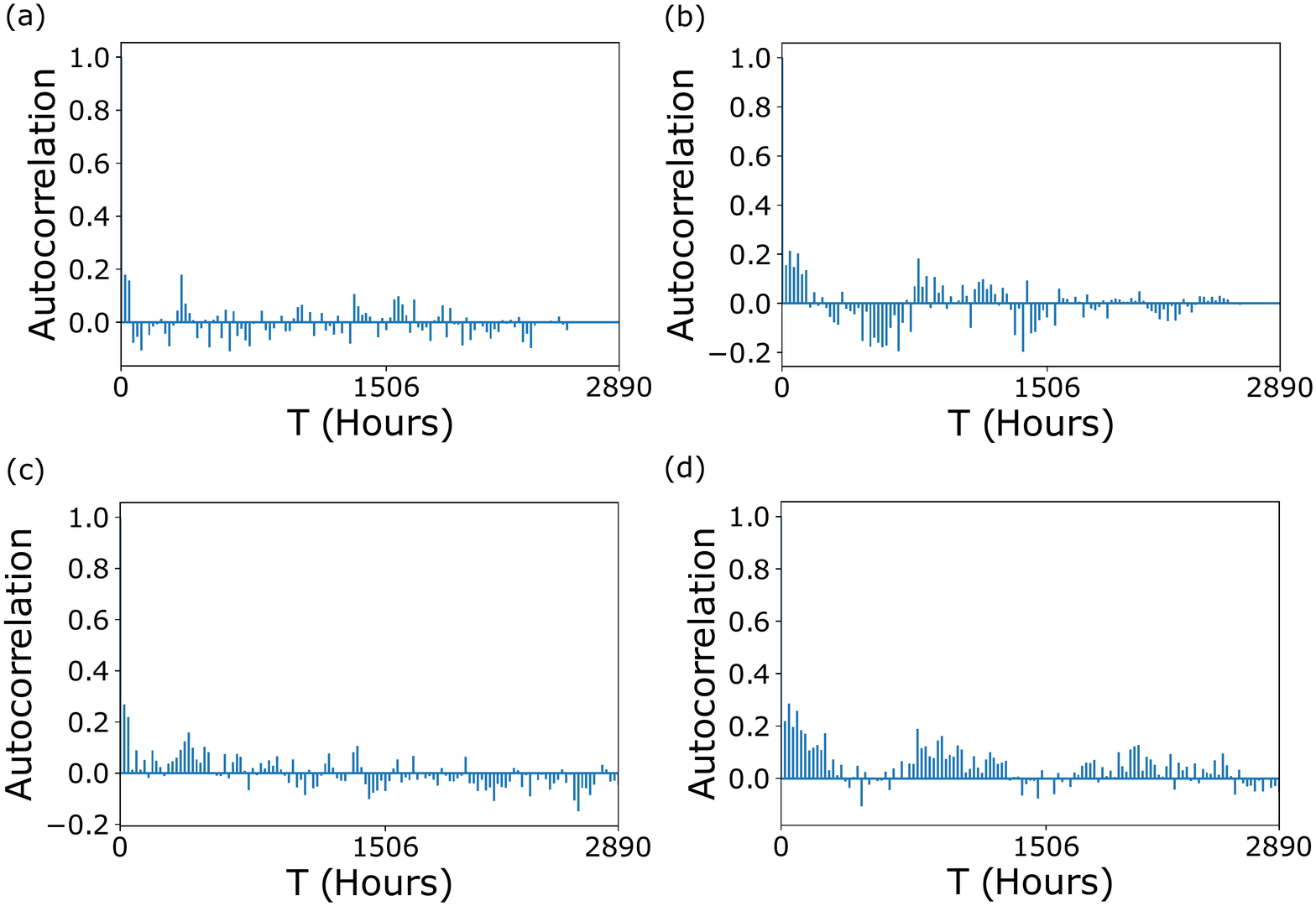}
	\caption{(a) and (b) autocorrelation as a function of measurement lag time for qubits 1 and 13, respectively, for an $\sim$2890 hour time period between temperature excursions. (c) and (d) are the autocorrelations using the entire time series, $\sim$2890 hours, for the corresponding qubits 1 and 13. The autocorrelation is detrended using the mean of the time series and normalized using the estimated variance.}
	\label{fig:autoWoutTexcursion}
\end{figure}

{\color{black} 
There were mixing chamber plate temperature excursions during the duration of the $\sim$9 month $T_1(T_i)$ measurement time series, Fig. \ref{fig:autocorrelation}. To address doubts about the impact of the temperature excursions, we show autocorrelation for representative qubits for the longest time series in which there are no thermal excursions for comparison, Fig \ref{fig:autoWoutTexcursion}. The qualitative behavior and magnitudes of the long time correlations are similar to those cases with the temperature excursions. Exact quantitative agreement does change but this would also be expected from truncating the time series at any different starting time. 

In Appendix \ref{autocorrelation} we show that the autocorrelation time is order of days. This is consistent with observing no strong effects of the temperature excursions on the weak stationarity of the time series or long term behavior of the distributions. That is, a temperature excursion would be expected to have a very localized effect on a $\sim$9 month time series behavior. The measurements using spectroscopy were done when the temperature was stable.

We conclude that the temperature excursions don't effect the conclusions of the paper.}

\section{Ergodicity of $T_1(\omega_q,T_i)$ time series}
\label{T1ergodicity}
{\color{black}
A foundational question is whether a $T_1$ time series behaves ergodically. That is, the time series $T_1(T_i)$ both converges to a $\langle T_1 \rangle$ (e.g., after sufficient time lag) and the pair correlations are well behaved (e.g., decay at long lag). Ergodicity is not guaranteed, drift or $1/f^{\alpha}$ behavior being illustrative reasons to doubt whether a reliable long time estimator of a physical property can be obtained. Since breaking of ergodicity can signal physical phenomena of interest such as switching between phases (i.e., isolated systems of an ensemble) with distinct mean values including special cases of spectral diffusion \cite{madzik_controllable_2020,brokmann_statistical_2003}, the establishment of whether the $T_1$ time series behaves ergodically represents an important step in clarifying the dynamics of the $T_1$ fluctuations.

The ergodic assumption is that given sufficient time a system will visit through all the accessible states (i.e., values) available to it. Such a sufficiently long time series trajectory can then be divided into $k$ independent subsets to form an ensemble of $k$ new systems that should represent the statistical behavior of the original system at any given time index, $m$ of the $k$ systems, Fig. \ref{fig:SubsetTimes} (a) \cite{Reif:1965uf}. The new time series of the systems are re-indexed, $m$, with equal lengths. An ensemble average is defined by selecting from the same time index for all equally sized subsets,

\begin{equation}
    \{T_1(T_m)\} = \frac{1}{k}\sum_{k}T_1(T_m)^{(k)} \mid m = constant
\end{equation}
and for an ergodic system,
\begin{equation}
    \{T_1(T_m)\} = \langle T_1 \rangle \simeq \langle T_1 \rangle_T
\end{equation}
when averaging across the ensemble of newly defined systems for any time index $i$. 
We test the $\sim$9 month $T_1$ time series for ergodic behavior. Autocorrelation of the different qubit time series shows some correlation over 1-2 days (i.e., first several measurements), see Appendix \ref{autocorrelation}. We examine a range of ensembles of size, $2 \leq k_{max} \leq 40$ containing $160 \geq m_{max} \geq 5$ time points, respectively. 

Ensemble averages are distributed around the time average $\langle T_1 \rangle_T$, Fig. \ref{fig:SubsetTimes} (c). We test the likelihood that the $\{T_1(T_m)\}$ is statistically indistinguishable from $\langle T_1 \rangle_T$ using a t-test comparison of the mean values of the two $T_1$ estimators. We show $Q_{13}$ results as an example of the analysis, Fig. \ref{fig:SubsetTimes} (d). In general the ensemble means are statistically indistinguishable from the $\sim$9 month means, $\{T_1\} \sim \langle T_1 \rangle_T$. Similar results are observed for the remaining odd qubits.   

\begin{figure*}[th]
	\includegraphics[width=\linewidth, height=45mm, clip,trim = 0.0mm 90.0mm 0.0mm 45.0mm]{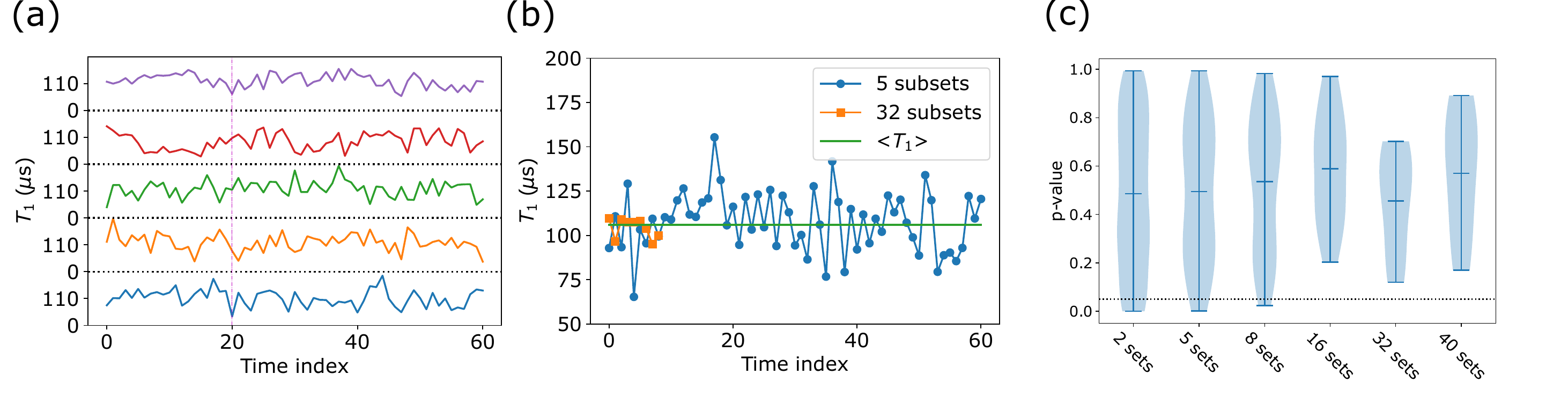}
	\caption{{\color{black}(a) The $k$ = 5 systems (i.e., subsets) formed from the $Q_{13}$ time series. The $T_1(T_i)$ measurements are re-indexed to form equal sized subsets with a new maximum index, $m$, of $m_{max}$ = 61. An ensemble, a set of 5 points in this case, is formed from the subsets for each $m$ time index. The vertical line is a guide to the eye representing one ensemble \{$T_{1,k}(T_m$ = $20)$\}. (b) Ensemble average dependence on time index for the 5 subset and 32 subset partitions. (c) violin plot results of t-tests for different system choices, $k_{max}$. More systems in the ensemble, $k_{max}$, corresponds to shorter time series. The dotted line marks 0.05 as a guide to the eye.}}
	\label{fig:SubsetTimes}
\end{figure*}

We note that we also investigated statistical tests of independence of the subsets. The overall conclusions are not changed when subsets are rejected from the ensemble based on failing statistical independence tests (i.e., the non-parametric Wald-Wolfowitz runs test\cite{Wald-Wolfowitz_10.2307/2235925}) as the majority of subsets are found independent to the limits of the sensitivity of the runs test.}

\section{Ergodicity of ensemble averaging of $T_1(\omega,t_i)$}
\label{T1fcErgodicity}
{\color{black}
We found in Appendix \ref{TimeseriesStationarity} that the $T_1$ time series of the qubits was generally weakly stationary as well as behaving ergodically, Appendix \ref{T1ergodicity}. The time series $T_1(\omega_q + \omega_j, t_i)$ can also be expected to behave as a weakly stationary ergodic time series as there is nothing uniquely distinctive about the frequency $\omega_q$. We further explicitly note that the sums and averages of neighboring time series will be weakly stationary and ergodic.

For frequencies close to $\omega_q$, we may assume $\lim_{\omega_j\to 0} \langle T_1(\omega_q + \omega_j) \rangle \rightarrow \langle T_1 \rangle_T$. We then ask whether an accurate $\langle T_1 \rangle_T$ estimator may be formed from an ensemble of neighboring $T_1(\omega_q + \omega_j, t_i)$ values. That is:

\begin{equation}
    \{T_1(t_i)\} = \frac{1}{S}\sum_{j=0}^{S}T_1(\omega_q +(j\chi-\Delta \omega),t_i) \simeq \langle T_1(\omega_q) \rangle_T
\end{equation}
where $S$ is the number of frequencies at which $T_1$ is sampled to form an ensemble estimator; $\chi$ is the frequency spacing of sampling in a single scan (see main text section \ref{process} or Fig. \ref{fig:FreqCorrs} inset); and $\Delta \omega$ is the maximum frequency range sampled, parameterized by $\chi$ as $\Delta \omega = \frac{(S-1)}{2}\chi$ (see section \ref{process} in main text).


To develop a heuristic for $\chi$, we choose a minimum frequency spacing $\omega_{j} - \omega_{j+1}$ that produces approximately independent $T_1(\omega_q + \omega_j,t_i)$s in the ensemble average of $\{T_1(t_i)\}$ (i.e., weak correlation with neighboring $T_1(\omega_q + \omega_{j\pm1},t_i)$). We calculate the correlation between frequencies to identify a $\chi$ that reduces the correlation below $\sim$0.2. We show an illustrative example of the autocorrelation of Stark shifted frequencies, $\langle T_1(\omega,0)T_1(\omega + \omega_{lag},0) \rangle $), for $Q_3$, Fig. \ref{fig:FreqCorrs}(a). The autocorrelation shows a fall off to weak correlation over $\sim$1-2 MHz centered around $\omega_{lag} \sim$ 0 for a single time index. Similar magnitude fall offs were observed for all qubits examined. This suggests a heuristic spacing of $\chi \simeq$ 1-2 MHz to obtain weakly correlated $T_1(\omega_q + \omega_j)$ with $T_1(\omega_q)$.



\begin{figure}[h]
	\includegraphics[width=75mm clip,trim = 40.0mm 20.0mm 0.0mm 0.0mm]{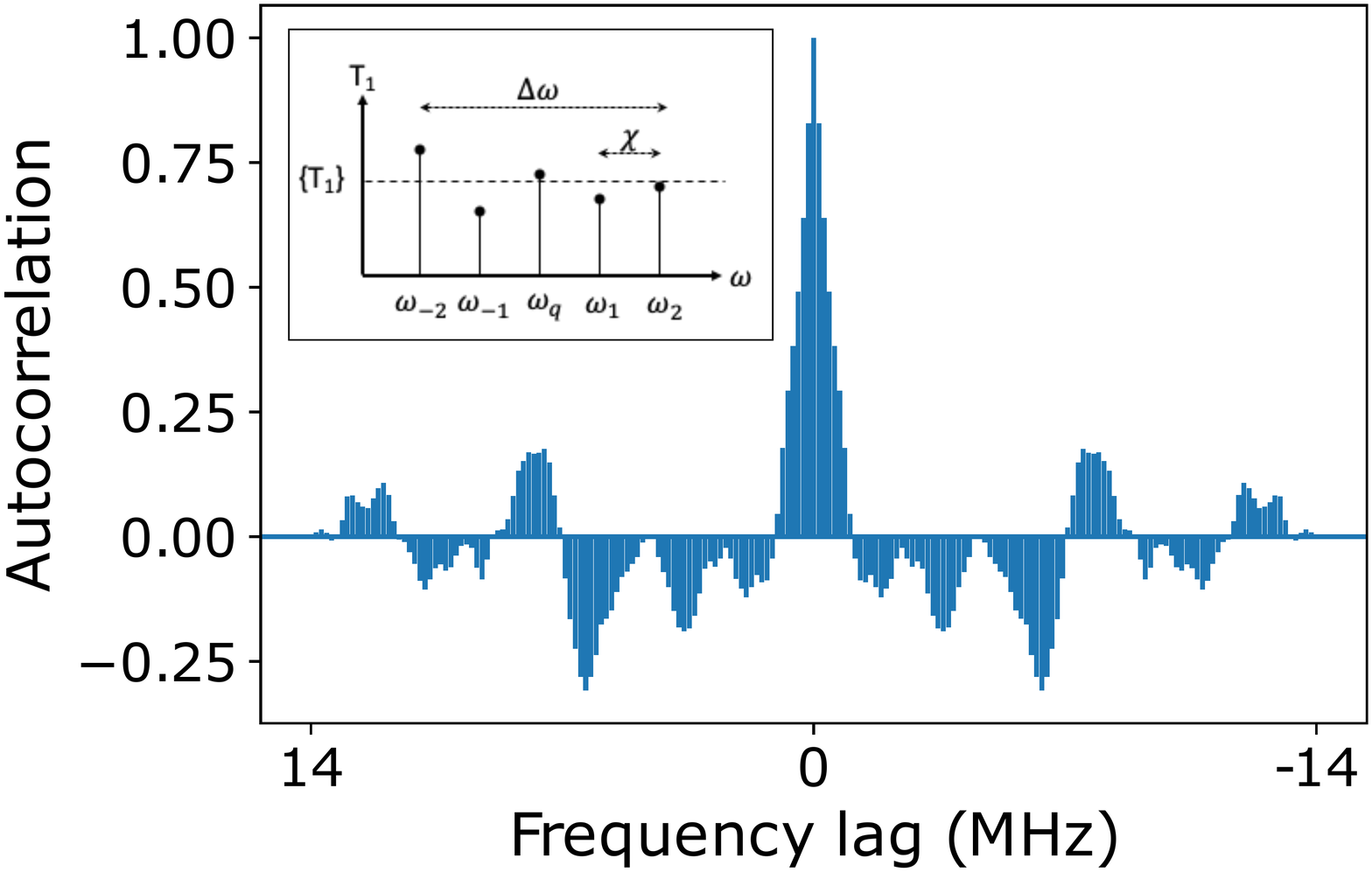}
	\caption{\color{black} (a) Autocorrelation of $Q_3$ at $t$ = 0. The autocorrelation is for negative Stark shift only. The autocorrelation is normalized and mean value detrended. (inset) Schematic of the formation of an ensemble average, $\{T_1\}$, from a set of $T_1(\omega_q + \omega_j,t_i)$ spaced by a frequency $\chi$ to minimize correlation between the $T_1$ values and sampling a range, $\Delta \omega$.}
	\label{fig:FreqCorrs}
\end{figure}

We empirically test the sensitivity of the error in the $\{T_1\}$ estimator to changing $\Delta \omega$ using $\chi$ = 2 MHz.  We define $\{T_1\}$ estimator error for each qubit as, $\frac{\mid \{T_1(t_i)\} - \langle T_1 \rangle_T \mid}{\langle T_1 \rangle_T}$. The error dependence, averaged over the 75 time steps for different $\Delta \omega$ (i.e., different $\chi$), is shown in Fig. \ref{fig:QuasiErgodic} (a). Notably the qubits have different dependencies on increasing $\Delta \omega$. There is no single $\Delta \omega$ that is optimal for every qubit. The differences between qubits helps explain the complex $\langle R(\Delta \omega) \rangle_{t_{0 \rightarrow n}}$ dependence shown in the main text, Fig. \ref{fig:Fig4} (c). The underlying cause for the differences is perhaps related to the details of the local spectral diffusion for each qubit. The nearly 1:1 relationship is further supported by measurements on other IBM devices, see Appendix \ref{OneToOne}. 

We now ask whether $\{T_1\}\simeq\langle T_1 \rangle_T$. The 75 $\{T_1\}$ distributions for each qubit are compared to the $\langle T_1 \rangle_T$ time series distributions using a t-test. Many of the 75 time indexed $\{T_1(t_i)\}$s are statistically indistinguishable but many are not. The probability of $\{T_1\}$ being indistinguishable from $\langle T_1 \rangle_T$ is shown in Fig. \ref{fig:QuasiErgodic} (b). 


We illustrate the proximity of the estimator by showing the combined distribution of the 75 means for $Q_7$, Fig. \ref{fig:QuasiErgodic} (c) over the $\sim$270 hours. The means, $\{T_1(t_i)\}$ and $\langle T_1 \rangle_T$, are within a few $\mu$s, less than a $\sigma$, but the null hypothesis (i.e., indistinguishable) is rejected because the distributions are sufficiently different. It is possible that the overall $\{T_1\}$ distribution would converge accumulating over a longer time. The $\sim$270 hours is insufficient for all the $T_1(\omega_q + \omega_j,t_i)$ time series to unambiguously converge around a mean value, see for example Fig. \ref{fig:SpecAutoCorr} (a) in Appendix \ref{Rspec}. Using longer periods of time to measure $\langle T_1(\omega_q + \omega_j) \rangle$ could therefore potentially lead $\{T_1\}$ to be more strictly ergodic. 

Empirically, $\Delta \omega \leq$ 10 MHz produces $\{T_1(t_i)\}$ distributions that are more likely to pass the t-test. The R values are also slightly better ranging from 0.89 to 0.91, reflecting that regardless of choice we find $\{T_1\} \simeq \langle T_1 \rangle_T \pm\sigma$. For illustration, we show the $\langle T_1 \rangle_T$, $\{T_1(t_i)\}$ pairs for $\Delta \omega$ = $\pm$6 MHz and $\chi$ = 2 MHz in Fig. \ref{fig:QuasiErgodic} (d), which has a Pearson R = 0.91. A very reliable 1:1 relationship is observed in many other qubits measured on other IBM devices, see Appendix \ref{OneToOne}.

\begin{figure}[h]
	\includegraphics[height=57mm, clip,trim = 0.0mm 35.0mm 0.0mm 0.0mm]{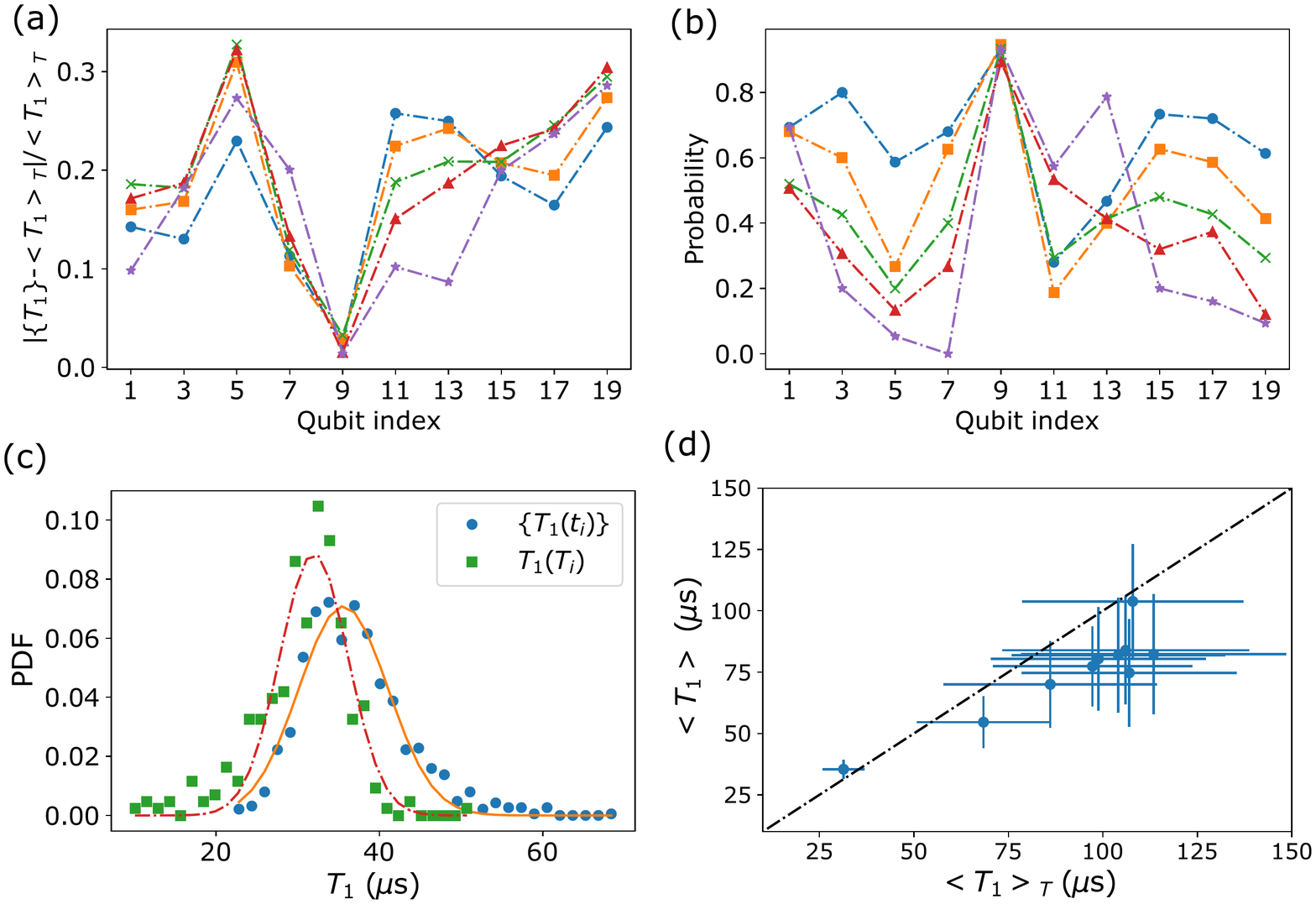}
	\caption{\color{black} (a) Dependence of mean estimator error on $\Delta \omega$ using a frequency spacing of $\chi$ = 2 MHz. The $\Delta \omega$ shown are $\pm$2 MHz ($\bullet$), $\pm$4 MHz ($\blacksquare$), $\pm$6 MHz ($\times$), $\pm$10 MHz ($\blacktriangle$) and $\pm$15 MHz ($\bigstar$). (b) Probability, out of 75 t-tests, that $\{T_1\}$ and $\langle T_1 \rangle_T$ are found indistinguishable as a function of range $\Delta \omega$ using a frequency step, $\chi$ = 2 MHz. The $\Delta$s are $\pm$2 MHz ($\bullet$), $\pm$4 MHz ($\blacksquare$), $\pm$6 MHz ($\times$), $\pm$10 MHz ($\blacktriangle$) and $\pm$15 MHz ($\bigstar$).(c) Histogram of $T_1$ measured at frequency steps of 2 MHz over $\Delta \omega$ = $\pm$15 MHz for $Q_7$. Overlaid are the $T_1(T_i)$ measurements from the $\sim$9 month time series. Normal fits for both distributions are shown in solid lines. (d) $\{T_1\}$ as a function of $<T_1>_T$ for each qubit. One sigma standard deviations of the distributions are shown as error bars for each $\{T_1\}$. $\Delta \omega$ = $\pm$6 MHz and $\chi$ = 2  MHz.}
	\label{fig:QuasiErgodic}
\end{figure}

To summarize, $\{T_1\} \simeq\langle T_1 \rangle_T \pm\sigma$. More statistically rigorous comparison by t-test indicates that $\{T_1(\Delta \omega,\chi)\}$ is more quasi-ergodic than strictly ergodic as it produces indistinguishable estimates of $\langle T_1 \rangle_T$ for less than 100\% of the ensembles. The quasi-ergodic results were found for $\chi$ and $\Delta \omega$ that were heuristically chosen and applied to all qubits. Individually optimized $\chi$ and $\Delta \omega$ will reduce disagreement and should become fully ergodic, certainly in the limit of $\Delta \omega, \chi \rightarrow 0$ converging trivially on the single time series $T_1(\omega_q,t_i)$. Optimal choices to achieve full ergodocity, while minimizing the number of measurements (e.g., total time to obtain the $T_1$ estimator), are left for future work. We  speculate that this might include forming an ensemble average with a physical model guided weighting of the $j^{th}$ elements of $\{T_1(\omega_q + \omega_j)\}$.  For immediate application of this approach, similar magnitude $\Delta \omega$ and $\chi$ values could be applied to other devices with the expectation that similar magnitude R values will be obtained as the R value is not strongly sensitive to the detailed choice of $\Delta \omega$ and $\chi$, see Appendix \ref{OneToOne}.    
}
\section{$R_{T_1}$ time dependence estimate from spectroscopy}
\label{Rspec}

\begin{figure*}[t!]
	\includegraphics[width=\textwidth,height=5cm, clip,trim = 0.0mm 85.0mm 0.0mm 0.0mm]{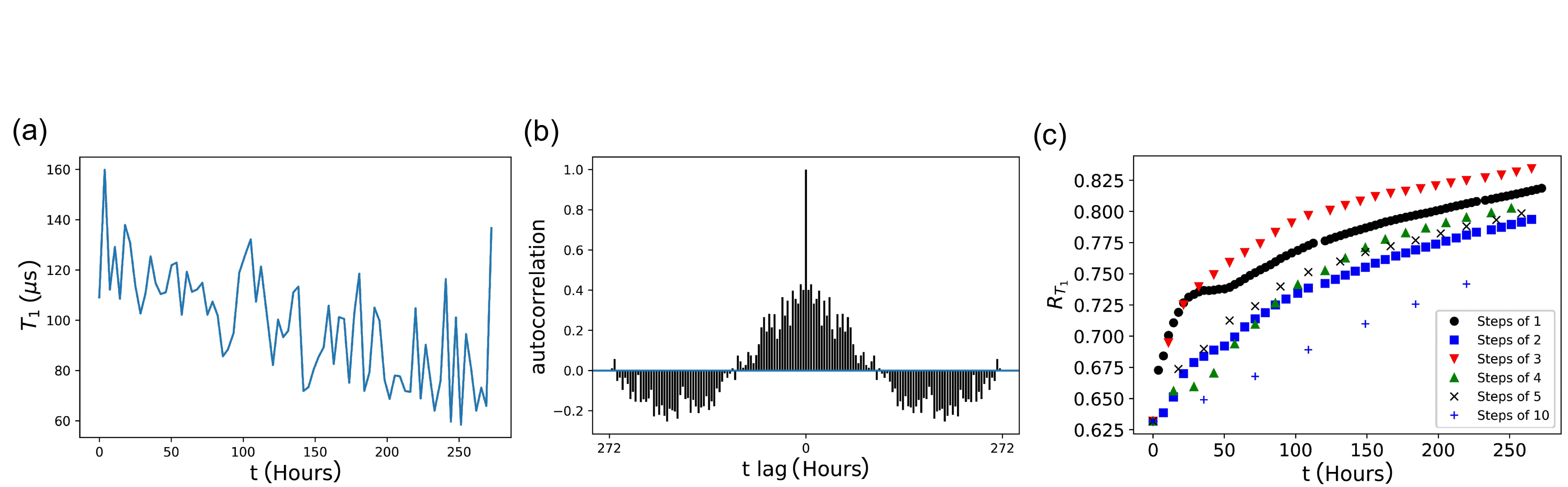}
	\caption{(a) $T_1$ time series calculated from $P_1$ spectroscopy for $Q_{15}$ at zero frequency shift, (b) autocorrelation of the $Q_{15}$ time series shown in (a) detrended with the mean and using padded zeros at the end of the time series to maintain a constant number of elements for all negative and positive lags (i.e., see online documentation for numpy.correlate). (c) Pearson $R$ correlation between the odd qubits $T_1$ from spectroscopy and $\langle T_1 \rangle_T$. The autocorrelation is done for every time points or intervals as indicated in the legend.}
	\label{fig:SpecAutoCorr}
\end{figure*}

We examine autocorrelation of $T_1$ from the data of Fig. \ref{fig:Figure4} and \ref{fig:FigE1} at $\Delta\omega_q$ = 0 MHz. The $T_1$ time series (see Fig. \ref{fig:SpecAutoCorr}(a)) for each of the qubits is deduced assuming,
\begin{equation}
   T_1 = \frac{-\tau}{\ln(P_1(\Delta\omega_q = 0))},
\end{equation}
where $\tau = 50\mu$s is the $T_1$ delay time, and $P_1 (\Delta\omega_q = 0)$ is the measured probability of being in the $| 1 \rangle$ state at time $\tau$ at the bare qubit frequency. 

Autocorrelation of the time series in many of the qubits show decaying correlation over the first 10-30 hours followed by weaker autocorrelation at longer times consistent with the longer $T_1$ series from the $\sim$9 month data set. This is seen in Fig. \ref{fig:SpecAutoCorr}(b). The autocorrelation is detrended with the mean and normalized to its estimated variance. Some qubits appear to have longer term drifts, see appendix \ref{autocorrelation}.

We calculate the Pearson $R$ correlations to the odd qubit $\langle T_1 \rangle_T$ in Fig. \ref{fig:SpecAutoCorr} (c). The correlations are shown for time series for different intervals of time available in the data set. This is to highlight the effects of correlation. For example, if the interval in time was doubled in an uncorrelated time series, it will double the time to achieve the same $R$ value on average. However, if there are strong correlations, increasing the interval times can decrease the time needed to achieve the same $R$ value.
Achieving $R$ $\sim$ 0.8 requires order of days or longer if the best interval is unknown. We note that the dependence on interval does vary some depending on what frequency is used for the $P_1$ data. The autocorrelation hampers achieving strong correlation (i.e., $R$ $\sim$ 0.8) in times shorter than 1-2 days.  

\section{Time dependence of TLS linewidths}
\label{line widths}
The time dependence of the TLS position in frequency is a topic of great interest \cite{muller_towards_2019,klimov_fluctuations_2018,schlor2019,deGraafeabc5055}. The linewidth (i.e., the width of the distribution of TLS frequency positions over time) is, for example, suggested to depend on the volume density of thermal fluctuators (TF) \cite{black_spectral_1977,klimov_fluctuations_2018}. TFs are low energy TLSs (i.e., $E_{TLS} \lesssim kT$). They are called thermal fluctuators (TF) because their configurations dynamically change in time due to thermal excitation. This results in a bath coupling to the higher energy TLSs (e.g., $E_{TLS} \sim \hbar\omega_q$) that produces the TLS spectral diffusion. A linewidth time dependence is, therefore, expected to depend on the TF density and coupling strength. An example of this dependence has been simulated in previous work \cite{klimov_fluctuations_2018}. This appendix provides a reference and comparison to previous TLS linewidth characterization. We present TLS linewidths for $Q_{15}$. 

\begin{figure}[th]
    \centering
    \includegraphics[width=85mm, clip,trim = 0.0mm 3.0mm 0.0mm 7.0mm]{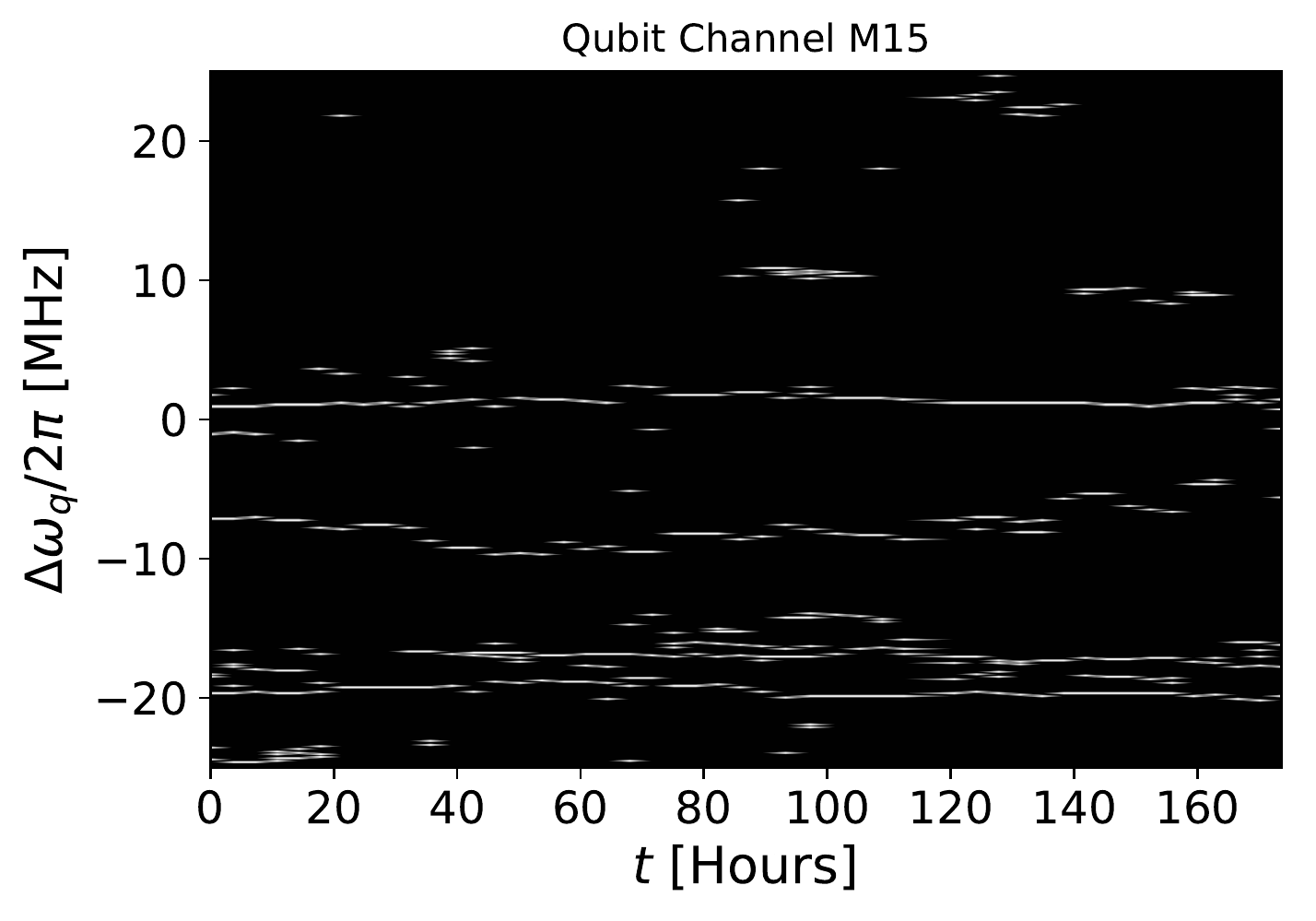}
    \caption{Peak locations of TLSs using a binary filtered replot of $Q_{15}$ with a $P_1$ = 0.315 threshold at a delay of $50\mu$s.}
    \label{fig:PeakSpectroscopy}
\end{figure}

The spectroscopy produces a discrete function of $P_1$ bins, $P_1(\omega,\tau$,t). To track the time evolution of the TLS position, we putatively associate each minima of a significant dip in $P_1$ with a TLS. We find the position of each min($P_1(\omega,\tau$,t)) and record the location for each time slice. We only record the TLS position if min($P_1)$ is below a threshold, $P_{th}$, of 0.315 to remove spurious markers of TLS location due to smaller fluctuations in $P_1$ (i.e., we focus on strongly coupled TLSs). The threshold of $P_{\textrm{th}} = $0.315 corresponds to a $T_1 < 43 \mu$s. The threshold was chosen by visual inspection to best minimize spurious points. The resulting TLS tracks are shown in Fig. \ref{fig:PeakSpectroscopy}. White indicates a min($P_1(\omega,\tau$,t)). As mentioned earlier in the paper, similar qualitative behavior is observed in the other qubits, see for example spectroscopy of odd qubits in appendix \ref{other spectroscopy} as well as in other works ~\cite{muller_towards_2019,klimov_fluctuations_2018}.

\begin{figure}[th]
	\includegraphics[width=85mm, clip,trim = 0.0mm 2.0mm 0.0mm 8mm]{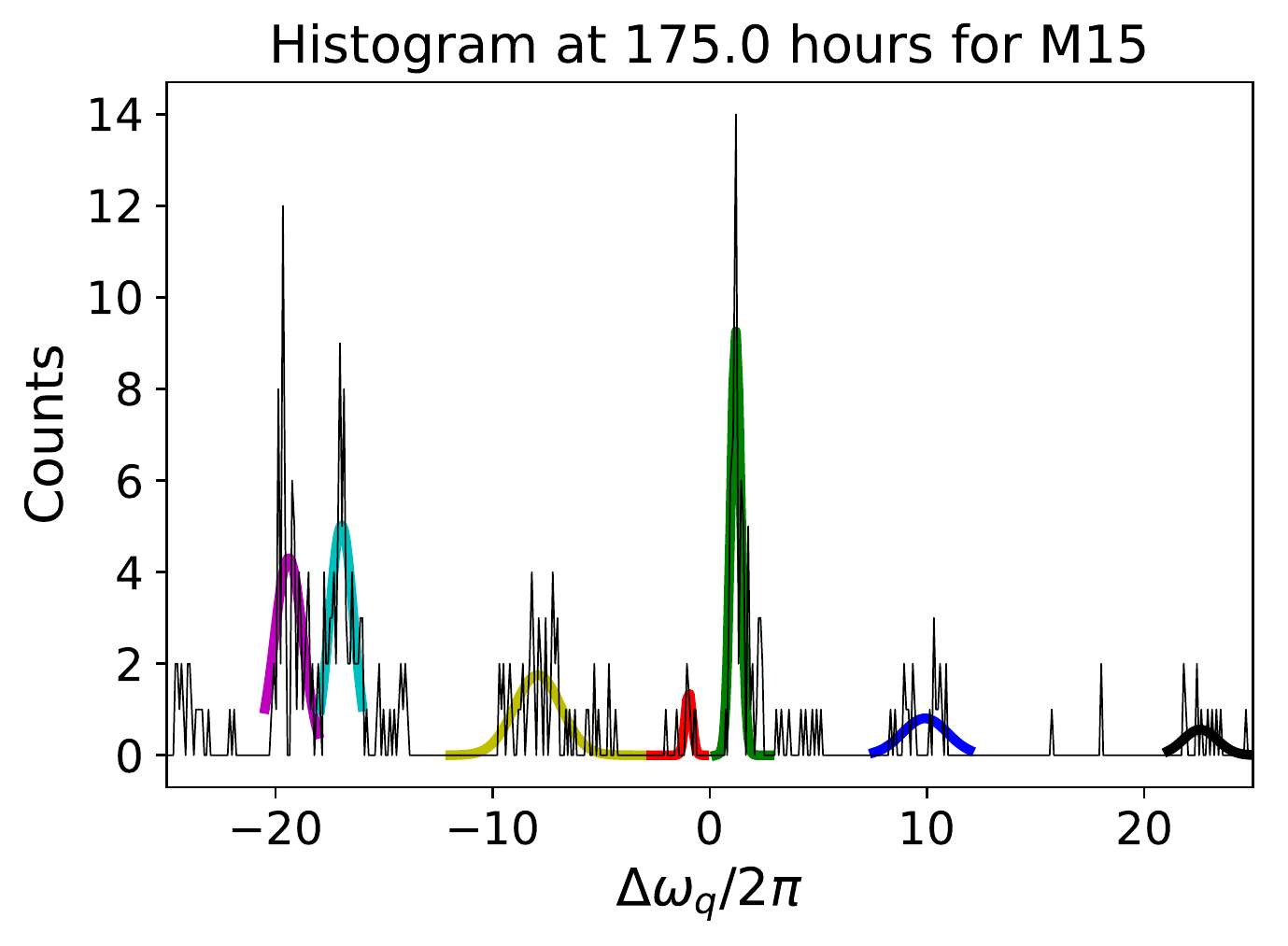}
	\caption{(a) histogram of the TLS positions from Fig. \ref{fig:PeakSpectroscopy}. The distribution of TLS positions are fit to Gaussians. {\color{black}The range over which the fit was done is indicated by the truncation points of the Guassians.}}
	\label{fig:Histogram}
\end{figure}

A linewidth can be crudely estimated by a cumulative histogram of the TLS positions as a function of time. The resulting histogram for $Q_{15}$ is shown in Fig. \ref{fig:Histogram}. Previous reports have fit the time dependence of spectral diffusion linewidths with a normal distribution ~\cite{klimov_fluctuations_2018,herzog_transient_1956} although we note that other functional forms are also predicted ~\cite{black_spectral_1977}. Nevertheless, to obtain a simple quantitative description of the linewidth and time dependence, we follow the phenomenological model of a normally distributed peak position, fitting each linewidth with a Gaussian and reporting the standard deviations, $\sigma$, in Table \ref{table:difftable}. We report diffusivities $D_{1d}$ assuming a standard one-dimensional model of the following form:

\begin{equation}
    C(t) = Ne^{{((\Delta \omega_q - \mu)^2}/{4\sigma(t)^2})}
\label{Gaussian}
\end{equation}
where $C$ is counts, $t$ is time, $\sigma$ is the standard deviation of the distribution, $\mu$ is the center frequency, $N$ is a fitting constant and for one-dimensional diffusion $\sigma(t) = \sqrt{2D_{1d}t}$. The spectral diffusion visually appears to be of the same order of magnitude in $Q_{15}$ as the other qubits so we present the values in Table \ref{table:difftable} as representative of the order of magnitude of Ds for the device. We also calculate $D_{K}$ with the units used by the recent work of Klimov et al. \cite{klimov_fluctuations_2018} (i.e., $\sigma(t) = 2D_Kt^{0.5}$) to provide rapid comparison to the extracted diffusivity and modeling in that work. However, we note that while we associate diffusivities to individual features that we track over time, the estimate provided in ref.~\cite{klimov_fluctuations_2018} is a single value fit to the consolidated linewidths of thirteen TLSs, an effective ensemble diffusivity.

Simulations of TLS spectral diffusion from Klimov et al. offer a suggestive and appealing link between TF volume densities and $D_K$ \cite{klimov_fluctuations_2018}. Despite the differences in analysis, the $D_{K}$s from this work for the most diffusive TLSs are over an order of magnitude smaller than reported by Klimov et al., 2.5 MHz (hr)$^{0.5}$, {\color{black}which might be interpreted as lower thermal fluctuator densities. However, the authors emphasize that this is a dubious conjecture. Doubts include: the model assumption of a time dependence of $\propto \sqrt{t}$; the related model assumption of an unbounded random walk of the TLS; and differences in the details of the temperature, between the two experimental setups, which might lead to differing numbers of active fluctuators.} Different time dependencies might indeed be expected (e.g., for different time regimes or dominant bath couplings \cite{herzog_transient_1956,black_spectral_1977,Witzel_PhysRevB.86.035452,madzik_controllable_2020}). Perhaps more importantly, the linewidths are likely truncated due to distance attenuated coupling mechanisms in the bath \cite{Klauder_PhysRev.125.912} (e.g., dipole coupling to thermal fluctuators). The assumption of ever increasing $\sigma(t)$ can lead to significant disagreement (i.e., for cases of longer time intervals of collection such as done in this work). We therefore do not presently put substantial weight on the comparison of extracted diffusivities until a more complete understanding of the linewidth time dependence is established. 
\begin{table}[h!]
\begin{tabular}{||c |c |c |c||} 
\hline
\makecell{Position \\ $(\textrm{MHz})$} & \makecell{$\sigma$ \\ $(\textrm{MHz}$)} &\makecell{$D_{K}$ \\ $(\textrm{MHz~hr}^{-0.5}$)} &\makecell{$D_{1d}$ \\ ($\textrm{MHz$^{2}$~hr}^{-1}$)} \\ 
\hline\hline
-19.3 & 0.65 & $2.4\times 10^{-2}$  & $1.1\times 10^{-3}$ \\
\hline
-16.9 & 0.55 & $2.6\times 10^{-2}$  & $0.8\times 10^{-4}$ \\
\hline
-7.9 & 1.06 & $3.9\times 10^{-2}$ & $3.1\times 10^{-3}$ \\ 
\hline
-0.97 & 0.16 & $6.1\times 10^{-3}$  & $7.3\times 10^{-5}$ \\
\hline
1.2 & 0.23 & $8.6\times 10^{-3}$  & $1.4\times 10^{-4}$ \\
\hline
9.9 & 1.03 & $3.1\times 10^{-3}$  & $3.0\times 10^{-3}$ \\
\hline
22.6 & 0.78 & $3.0\times 10^{-2}$  & $1.7\times 10^{-3}$ \\
\hline
\end{tabular}
\caption{Table \ref{table:difftable} Estimated standard deviations for each of the TLS peak distributions after 175 hours and diffusivities following  ref. \cite{klimov_fluctuations_2018} $D_K$ or a one-dimensional diffusivity $D_{1d}$.}
\label{table:difftable}
\end{table}

We also note two additional challenges to the accuracy of linewidth analysis of individual TLSs, beyond the limits of validity of the one-dimensional model, those being: (1) dips can overlap in ambiguous ways and (2) there can be uncertainty in assignment of positions related to other TLS-like features that migrate through the frequency region of interest. In Fig. \ref{fig:PeakSpectroscopy} one can see at least one other weaker dip that weaves between the two prominent ones in the -15 to -20 MHz range, for example.

\section{Almaden measurement details}
\label{Mdetails}
{\color{black}
The Almaden device was a deployed system with cloud access. A daily calibration was done, which included $T_1$ measurement. A database recorded calibration measurements and the measurement times. Some additional measurements were added to the database due to custom checks and recalibration of qubits that were outside of the daily cals. The database was queried from 2019-09-13 to 2020-07-15. 

The $T_1$ measurement was done for 41 time points logarithmically spaced up to 500 $\mu$s using 300 shots per time point. A simple exponential fit was made to the decay. The TLS spectroscopy was done using 501 frequency points per direction of Stark shift with 1000 repeated rounds for each point. The repetition time was 1 ms. This time can be substantially reduced with faster reset of the initial state, for example.
}

\section{Supporting evidence for 1:1 correlation between $\langle T_1 \rangle_{\omega,t}$ and $\langle T_1 \rangle_T$}
\label{OneToOne}
{\color{black}
We provide additional evidence that $\langle T_1 \rangle_{\omega,t}$ is a 1:1 estimator of $\langle T_1 \rangle_T$. We compile long time averages $\langle T_1 \rangle_T$ for 458 qubits and compare them to a single $T_1$ measurement in Figure \ref{fig:OneToOne} (a) to provide an illustrative example of statistical spread and resulting R value. A 1:1 guide to the eye is overlaid and a Pearson R of 0.72 is measured for the single $T_1$ estimator of $\langle T_1 \rangle_T$ for this instance. The length of the time series of daily $T_1$ measurements depends on the amount of time the device was deployed. The total time duration over which the $T_1$ measurements were done are indicated in Table \ref{table:DeployOnetoOne}. 

We contrast this with a $\langle T_1(t_i,n = 1) \rangle_{\omega,t}$ measured from a single spectroscopy scan of each of the qubits, Fig. \ref{fig:OneToOne} (b) randomly selected in the spectroscopy time series of measurements taken approximately every 6 hours. For the spectroscopy measurements, the $\omega_s = \pm$80 MHz and drive amplitude was swept to a fixed amplitude in both the negative and positive shifts. This results in a total $\Delta \omega \sim$ 25 MHz. Each qubit shifts slightly differently due to differences such as line attenuation. The R value between $\langle T_1(t_i) \rangle_{\omega}$ and $\langle T_1 \rangle_T$ was 0.82. Visually, a tighter concentration around a one-to-one correspondence with $\langle T_1 \rangle_T$ is observed from the $\langle T_1 \rangle_{\omega,t}$ estimator than relying on single $T_1$ measurements, consistent with observations made in the main text.

We also examine how $\langle T_1(n) \rangle_{\omega,t}$ converges with $\langle T_1 \rangle_T$ averaging over the duration of available $\sim$6 hour repeated spectroscopy measurements for each device. The time series durations for the spectroscopy, $t_{max}$, are indicated for each device in table \ref{table:DeployOnetoOne}. The R value improves to 0.91. The source of residual disagreement likely comes in part from the lack of custom optimization of $\Delta \omega$ and best choice of weighting of $T_1(\omega_q + \omega_j,t_i)$.

\begin{figure*}[!t]
	\includegraphics[width=\linewidth, height=7 cm, clip,trim = 1.0mm 80.0mm 0.0mm 25.0mm]{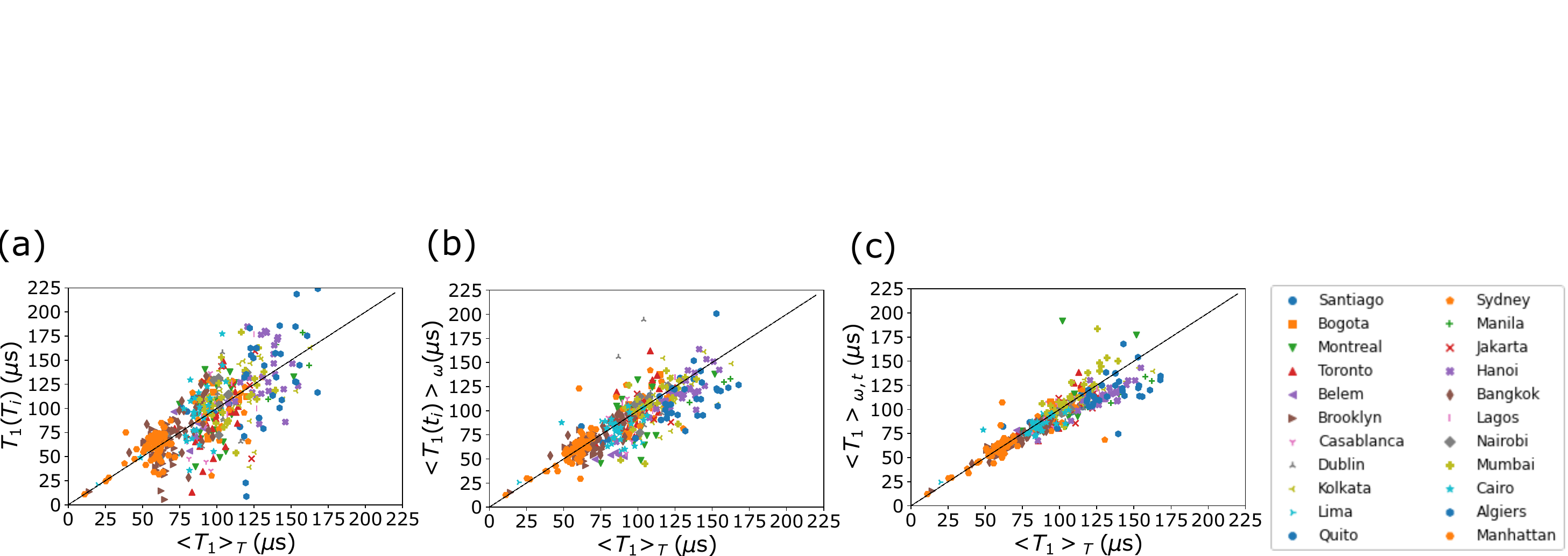}
	\caption{{\color{black}Scatter plots of $\langle T_1 \rangle_T$ compared to different estimators for 458 qubits from devices listed in table \ref{table:DeployOnetoOne}. (a) Comparison of $\langle T_1 \rangle_T$ to a single $T_1(T_i)$ measurement for each qubit randomly selected from the $T_1$ time series. (b) $\langle T_1(t_i) \rangle_{\omega}$ estimated from $P_1(\omega_q + \omega_j,t_i)$ measurements for a spectroscopy scan at a randomly selected time, $t_i$, in the spectroscopy time series. (c) $\langle T_1 \rangle_{\omega,t}$ averaging over all scans available, see the duration of the time series for each device in table \ref{table:DeployOnetoOne}. The Pearson Rs for the three cases are 0.72, 0.82 and 0.91, respectively}}
	\label{fig:OneToOne}
\end{figure*}

}
\begin{table}[h!]
\begin{tabular}{||c |c |c |c||} 
\hline
\makecell{Device} &\makecell{ Qubits} &\makecell{$T_{max}$ (days)} &\makecell{ $t_{max}$ (days) }\\
\hline\hline
Santiago & 5 & 453  & 169 \\
\hline
Bogota & 5 & 437  & 169 \\
\hline
Montreal & 27 & 452 & 220 \\ 
\hline
Toronto & 27 & 449  & 169 \\
\hline
Belem & 5 & 220 & 192 \\
\hline
Brooklyn & 65 & 145 & 108 \\
\hline
Casablana & 7 & 362  & 169 \\
\hline
Dublin & 27 & 350  & 168 \\
\hline
Kolkata & 27 & 236  & 180 \\
\hline
Lima & 5 & 216  & 192 \\
\hline
Quito & 5 & 222  & 192 \\
\hline
Sydney & 27 & 366  & 168 \\
\hline
Manila & 5 & 123 & 101 \\
\hline
Jakarta & 7 & 138 & 118 \\
\hline
Hanoi & 27 & 132 & 107 \\
\hline
Bangkok & 27 & 130 & 78 \\
\hline
Lagos & 7 & 68 & 45 \\
\hline
Nairobi & 7 & 76 & 55 \\
\hline
Mumbai & 27 & 278 & 198 \\
\hline
Cairo & 27 & 55 & 30 \\
\hline
Algiers & 27 & 73 & 51 \\
\hline
Manhattan & 65 & 402 & 191 \\
\hline
\end{tabular}
\caption{{\color{black} Device names, number of qubits and the length of time series for the $T_1$ measurements, $T_{max}$, and Stark spectroscopy, $t_{max}$. The $T_1$ measurements are taken every 24 hours and the spectroscopy measurements are taken approximately every 6 hours.}}
\label{table:DeployOnetoOne}
\end{table}

\end{document}